\documentclass[prl,twocolumn,superscriptaddress]{revtex4}

\usepackage{amsmath}
\usepackage{amssymb}
\usepackage{graphicx}
\usepackage{upgreek}
\usepackage[colorlinks=true, citecolor=blue, urlcolor=blue]{hyperref}

\newcommand{\tr}{{\rm Tr}}

\begin{document}
\title{Experimental investigation of uncertainty relations for non-Hermitian operators}
\author{Xinzhi Zhao}
\affiliation{School of Physical Science and Technology, Ningbo University, Ningbo, 315211, China}
\author{Xinglei Yu}
\affiliation{School of Physical Science and Technology, Ningbo University, Ningbo, 315211, China}
\author{Wenting Zhou}
\affiliation{School of Physical Science and Technology, Ningbo University, Ningbo, 315211, China}
\author{Chengjie Zhang}
\email{chengjie.zhang@gmail.com}
\affiliation{School of Physical Science and Technology, Ningbo University, Ningbo, 315211, China}
\affiliation{Hefei National Laboratory, University of Science and Technology of China, Hefei 230088, China}
\author{Jin-Shi Xu}
\email{jsxu@ustc.edu.cn}
\affiliation{CAS Key Laboratory of Quantum Information, University of Science and Technology of China, Hefei 230026, China}
\affiliation{CAS Center for Excellence in Quantum Information and Quantum Physics, University of Science and Technology of China, Hefei 230026, China}
\affiliation{Hefei National Laboratory, University of Science and Technology of China, Hefei 230088, China}
\author{Chuan-Feng Li}
\email{cfli@ustc.edu.cn}
\affiliation{CAS Key Laboratory of Quantum Information, University of Science and Technology of China, Hefei 230026, China}
\affiliation{CAS Center for Excellence in Quantum Information and Quantum Physics, University of Science and Technology of China, Hefei 230026, China}
\affiliation{Hefei National Laboratory, University of Science and Technology of China, Hefei 230088, China}
\author{Guang-Can Guo}
\affiliation{CAS Key Laboratory of Quantum Information, University of Science and Technology of China, Hefei 230026, China}
\affiliation{CAS Center for Excellence in Quantum Information and Quantum Physics, University of Science and Technology of China, Hefei 230026, China}
\affiliation{Hefei National Laboratory, University of Science and Technology of China, Hefei 230088, China}

\begin{abstract}
Uncertainty relations for Hermitian operators have been confirmed through many experiments. However, previous experiments have only tested the special case of non-Hermitian operators, i.e., uncertainty relations for unitary operators. In this study, we explore uncertainty relations for general non-Hermitian operators, which include Hermitian and unitary operators as special cases. We perform experiments with both real and complex non-Hermitian operators for qubit states, and confirm the validity of the uncertainty relations within the experimental error. Our results provide experimental evidence of uncertainty relations for non-Hermitian operators. Furthermore, our methods for realizing and measuring non-Hermitian operators are valuable in characterizing open-system dynamics and enhancing parameter estimation.
\end{abstract}

\maketitle

\textit{Introduction.---} Uncertainty relations are essential for the development of quantum physics, starting from the position-momentum uncertainty relation to Robertson-Schr\"{o}dinger uncertainty relation for any pair of observables \cite{Heis,Keen,Grup,Tson,Shro}. Uncertainty relations also have numerous important applications in quantum information theory, such as quantum random number generation \cite{Vall,Zhou}, entanglement detection \cite{Patr}, Einstein-Podolsky-Rosen steering \cite{Walb,Schn}, quantum metrology \cite{Giov}.

Quantum theory  represents physical observables as Hermitian operators that can be directly measured in experiments. However, not all operators in quantum theory are Hermitian. Some examples of non-Hermitian operators are  unitary operators, ladder operators, and effective  non-Hermitian Hamiltonians for open systems \cite{Zhan,Chen,Ahar,Rost}. Although it is not easy to experimentally measure non-Hermitian operators since they may have complex eigenvalues, non-Hermitian operators can still be measured in experiments by using certain methods, for instance measuring them via weak values \cite{Pati,Pati2}, decomposing any non-Hermitian operator into  complex-weighted sum of Hermitian operators \cite{Leach}. Non-Hermitian operators have many applications, including effective non-Hermitian Hamiltonian in quantum open systems \cite{eff1,eff2,eff4,eff5},  metamaterials \cite{memt1,memt2,memt3,memt4,memt5}, and nonreciprocal device \cite{nrd1,nrd2}, etc.

Various incompatible observables \cite{Hall,Macc,Shar,Mond,Rudn} and physical systems \cite{Erha,Suly,Spon,Licf,Prev,Roze,West,Ring,Kane} have been considered in the study and verification of uncertainty relations. A natural extension of this research is to consider uncertainty relations for non-Hermitian operators. On the one hand, several  theoretical results have been presented. Pati {\it et al.} have provided an uncertainty relation for two general non-Hermitian operators \cite{Pati}, and  for general unitary operators \cite{Bagc}.  Massar and Spindel have obtained uncertainty relation for the discrete fourier transform \cite{Mass}. Yu {\it et al.} have derived strong unitary uncertainty relations \cite{yu}. Moreover, uncertainty relations for $n$ general non-Hermitian operators have been shown in Ref. \cite{Zhao}. On the other hand, experimental results have also been reported. Strong unitary uncertainty relations have been investigated in experiments \cite {Tisc,Xiao,Wang}.  However, previous experiments have only tested  uncertainty relations for unitary operators, which are special cases of non-Hermitian operators. There is still a lack of experimental results on uncertainty relations for non-Hermitian operators that are not unitary.

In this work, we investigate the uncertainty relation for non-Hermitian operators. These operators often have complex eigenvalues that are difficult to measure experimentally. Thus, we need to devise a measurement method for non-Hermitian operators in the experiment. We consider real and complex non-Hermitian operators for single-qubit states, and derive the theoretical uncertainty relations for those non-Hermitian operators. We experimentally verify the uncertainty relations by using a Sagnac ring interferometer, and realize non-Hermitian operators by losing some photons. Our results provide experimental evidence of uncertainty relations for non-Hermitian operators. In addition, our experiment methods can be applied in a number of quantum areas, such as quantum parameter estimation, pseudo-Hermitian system and entanglement detection.

\begin{figure*}
\includegraphics[width=14cm]{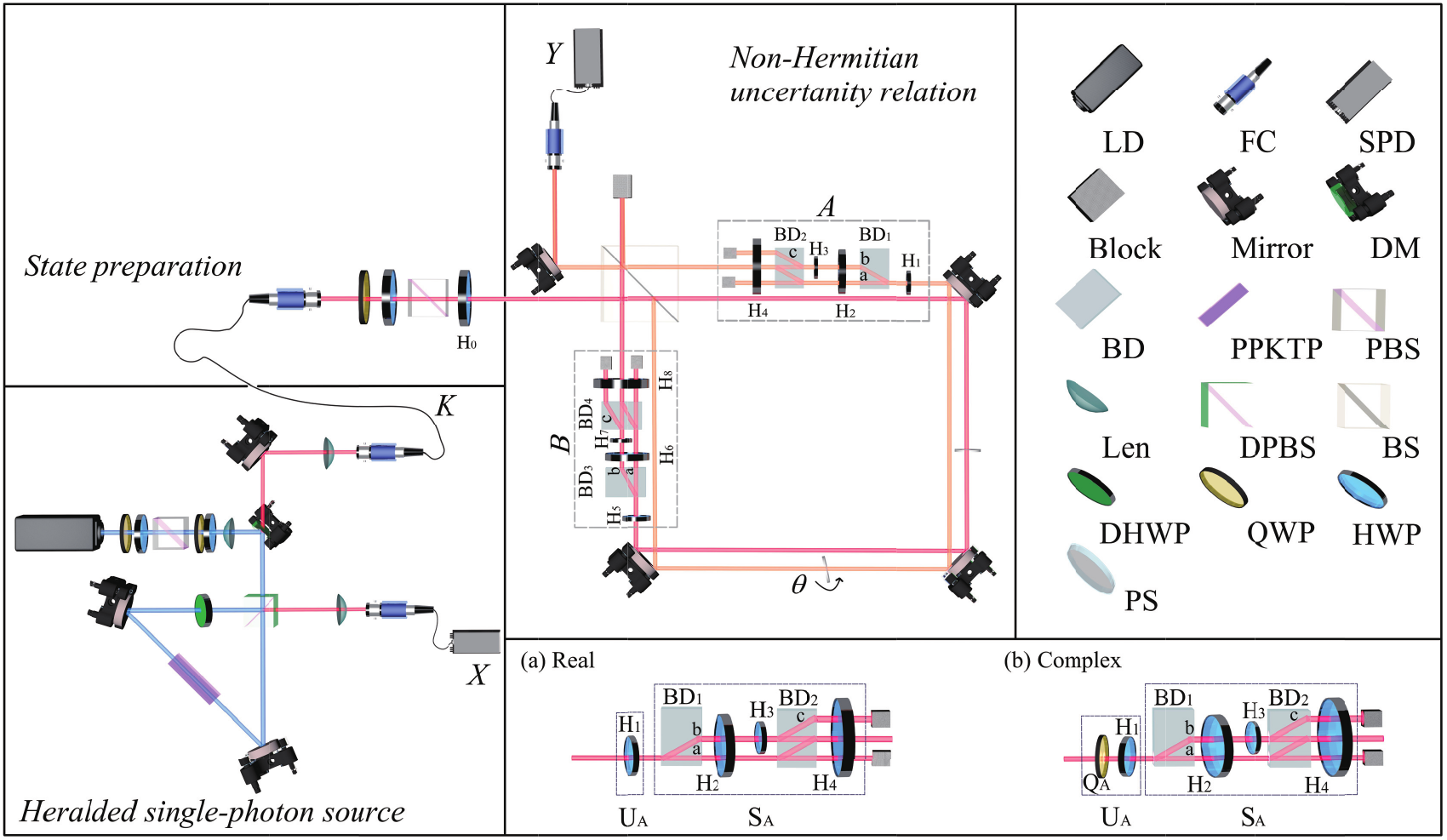}
\caption{An experimental setup for testing the uncertainty relations between two non-Hermitian operators $A$ and $B$ in two scenarios: case (a) both $A$ and $B$ are real, and case (b) both $A$ and $B$ are complex. Pairs of single photons are generated via SPDC using a type-II PPKTP crystal. A wave plate group consisting of a half wave plate (HWP), a quarter wave plate (QWP), a polarization beam splitter (PBS) and a HWP are used to prepare initial input states encoded in the polarization of signal photons.
We use a $50:50$ non-polarizing beam splitter (BS) to construct a shifted Sagnac ring interferometer, with small interferometers using two beam displacer (BD) crystals embedded in both transmission (red) and reflection (orange) paths.
Corresponding to the real case and the complex case, the non-Hermitian operator $A$ is realized by a small interferometer with HWPs and QWPs as shown in (a) and (b). The detailed construction of $A$ and $B$ can be found in Supplemental Material \cite{Supp}. The flat window on one path is tilted to act as a phase shifter (PS), while the fixed flat window on the other path keeps the path length difference within the coherent length. Two single photon detectors (SPDs) detect signal and idle photons, and coincidence counts are measured by a time-correlated single photon counting system. Dichroic mirror (DM); Dichroic  polarization beam splitter (DPBS); Dichroic half wave plate (DHWP). } \label{Fig1}
\end{figure*}

\textit{Uncertainty relations for non-Hermitian operators.---} The definition of variance of a non-Hermitian operator $O$ under a state $\varrho$ can be written as \cite{Pati,Anna} $\langle(\Delta O)^2\rangle:=\langle O^\dag O\rangle-\langle O^\dag \rangle \langle O \rangle$, where $\langle O \rangle=\tr(\varrho O)$. The first proof of the uncertainty relation for two non-Hermitian operators appeared  in Ref. \cite{Pati},
\begin{eqnarray}
\langle(\Delta A)^2\rangle \langle (\Delta B)^2\rangle \geq |\langle A^\dag B\rangle-\langle A^\dag\rangle \langle B\rangle|^2.\label{UR}
\end{eqnarray}
For any $n$ non-Hermitian operators $\{A_i\}_{i=1}^{n}$ in
$d$-dimensional Hilbert space, the uncertainty relation is represented by the matrix $M$, which is positive semidefinite \cite{Zhao}.
Let us consider two non-Hermitian operators $A$ and $B$ in a single qubit system. For an arbitrary pure state $|\varphi\rangle$,
we can define $|\phi_{1}\rangle :=|\varphi\rangle$, $|\phi_{2}\rangle :=A|\varphi\rangle$, $|\phi_{3}\rangle:=B|\varphi\rangle$, and $T_{ij}:=\langle\phi_{i}|\phi_{j}\rangle$  with $i,j=1,2,3$. From uncertain relation~(\ref{UR}), one can derive a simple uncertainty relation for these non-Hermitian operators in a single qubit system,
\begin{eqnarray}
\cfrac{|T_{12}|^2}{|T_{22}|}+\cfrac{|T_{31}|^2}{|T_{33}|}+\cfrac{|T_{23}|^2}{|T_{22} T_{33}|}-2\cfrac{|T_{23} T_{12} T_{31}|}{|T_{22} T_{33}|}\leq 1,\label{UR5}
\end{eqnarray}
details are given in Supplemental Material \cite{Supp}.

\textit{Experimental setup.---} As shown in Fig.~\ref{Fig1}, the main components of our device are a heralded single-photon source and a phase-adjustable Sagnac ring interferometer. The heralded single-photon source consists of a $405$ $\mathrm{nm}$ continuous-wave laser diode (LD) and a triangular Sagnac interferometer \cite{Baos,Teah}. In order to satisfy the heralded single-photon source requirements of our experiments, we selected a periodically poled ${\rm {KTiOPO_{4}}}$ (PPKTP) crystal with a size of $1\times 2\times 20$ mm, a poling period of $10.025$ $\upmu$m and the operating temperature was controlled at $25^{\circ}$C. Degenerate photon pairs produced by collinear type-II spontaneous parameter down-conversion (SPDC) are collected into the single mode fibers by the fiber couplers (FCs) after narrow-band filters. Since photons come in pairs, each coincidental detection of an idle photon indicates the presence of a signal photon, where idle and signal photons are counted by the single photon detectors (SPDs) X and Y, respectively. For more information about construction of the heralded single-photon source, see Supplemental Material \cite{Supp}.

In order to verify Eq.~(\ref{UR5}), we perform non-Hermitian operations by manipulating single-photon states with the  phase-adjustable Sagnac ring interferometer and  BD crystals. As illustrated in Fig.~\ref{Fig1}, $A$ and $B$ are the reflection  and transmission paths, respectively. If $A$ and $B$ are unitary operators, one can determine the value of $|\langle A^\dag B\rangle|$ for an input state $|\varphi\rangle$ by first noting that the average number of output photons is $N(A^\dag,B)=(\langle A^\dag A\rangle +\langle B^\dag B\rangle +2\cos{\theta}|\langle A^\dag B\rangle|)/4$, where $\theta$ is the phase difference between the two arms. Therefore, by varying  $\theta$  one has $N_{max}(A^\dag, A)=|\langle A^\dag A\rangle|$, $N_{max}(A^\dag, B)+N_{min}(A^\dag, B)=(|\langle A^\dag A\rangle|+|\langle B^\dag B\rangle|)/2$, and $N_{max}(A^\dag, B)-N_{min}(A^\dag, B)=|\langle A^\dag B\rangle|$, etc. The interference visibility $\gamma(A^\dag, B):=(N_{max}(A^\dag, B)-N_{min}(A^\dag, B))/(N_{max}(A^\dag, B)+N_{min}(A^\dag, B))=2|\langle A^\dag B\rangle|/(|\langle A^\dag A\rangle|+|\langle B^\dag B\rangle|)$ can be calculated from the number of photons \cite{Tisc}.

\begin{figure}
\begin{center}
\includegraphics[scale=0.20]{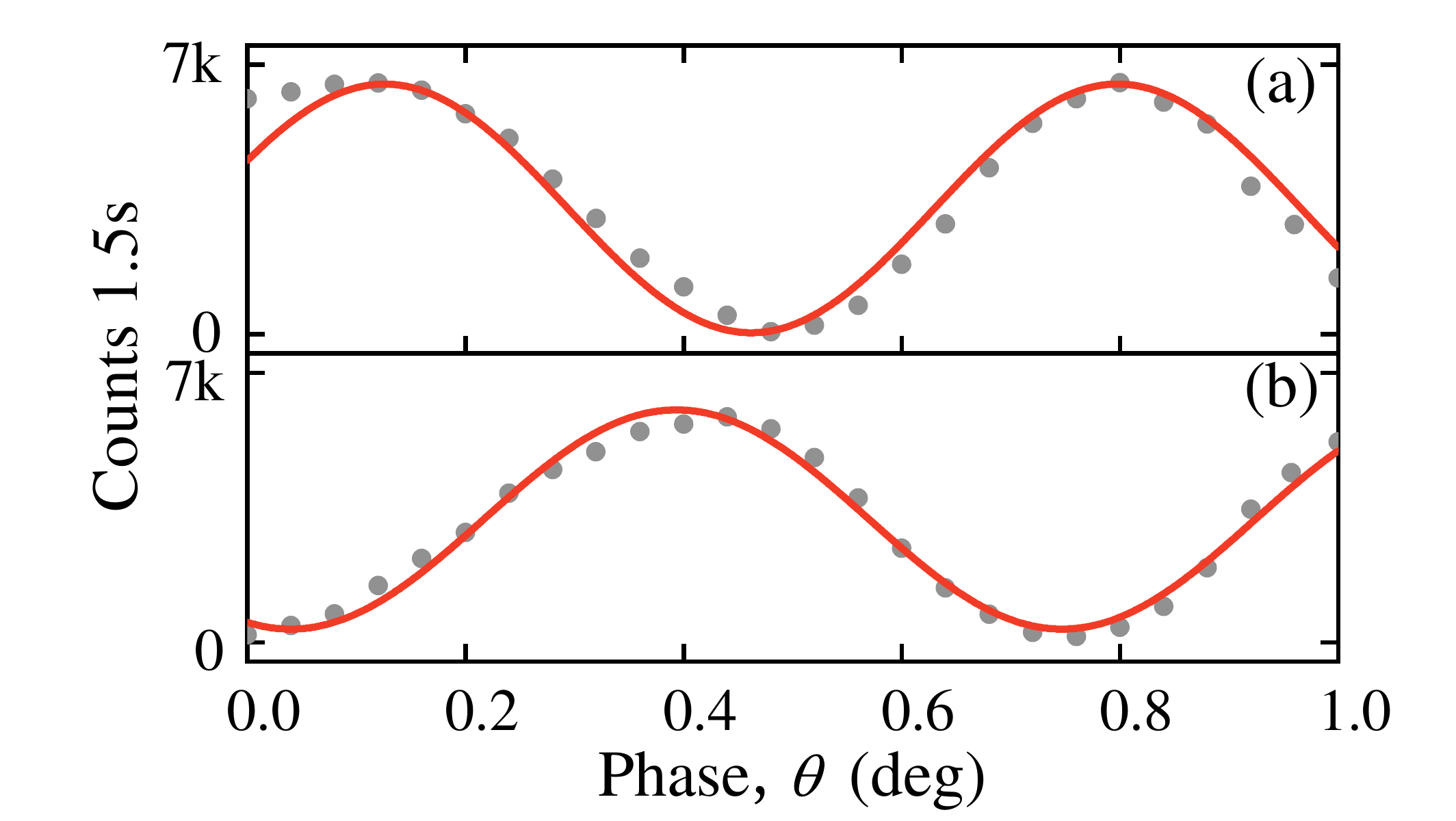}
\caption{Interferogram recorded by a coincidence counter. Non-Hermitian and identity operators are applied to each arm of the interference ring as follows (a) Transmission: $I$, Reflection: $I$. (b) Transmission: $B$, Reflection: $I$.}\label{Fig2}
\end{center}
\end{figure}

However, when $A$ and $B$ are general non-Hermitian operators rather than unitary operators, in order to realize the loss for non-Hermitian operators, we use two BD crystals nested on each path of the Sagnac interferometer.
Because some photons are lost, but the normalization must be guaranteed, we use an analogue of the interference visibility. In other words,  we measure the maximal and minimal values when both $A$ and $B$ are identity operators, and the sum of these values gives the total number of input photons. For example, $|T_{23}|=|\langle A^\dag B\rangle|$ can be written as
\begin{eqnarray}
|T_{23}|=\cfrac{N_{max}(A^\dag,B)-N_{min}(A^\dag,B)}{N_{max}(I,I)+N_{min}(I,I)}, 
\end{eqnarray}
where $N_{max}(I,I)+N_{min}(I,I)$ is the total number of input photons, details are given in Supplemental Material \cite{Supp}. To measure $N_{max}(A^\dag,B)$ and $N_{min}(A^\dag,B)$, we need to realize $A$ and $B$ in the reflection and transmission paths of the Sagnac interferometer in Fig.~\ref{Fig1}, respectively, and record the maximal and minimal coincidence counts. Similarly, when identity operators are prepared in both reflection and transmission paths, one can measure $N_{max}(I,I)$ and $N_{min}(I,I)$ via  the maximal and minimal coincidence counts, as shown in Fig.~\ref{Fig2}(a).
The values of other $|T_{ij}|$ are determined in a similar way as $|T_{23}|$. In order to experimentally verify Eq.~(\ref{UR5}), we need to measure $|T_{22}|$, $|T_{33}|$, $|T_{23}|$, $|T_{12}|$ and $|T_{31}|$ in the experiment. Moreover, we do the polar decompositions for the non-Hermitian operators $A=S_AU_A$ and $B=S_BU_B$ to realize them experimentally.

\begin{figure}
\begin{center}
\includegraphics[scale=0.20]{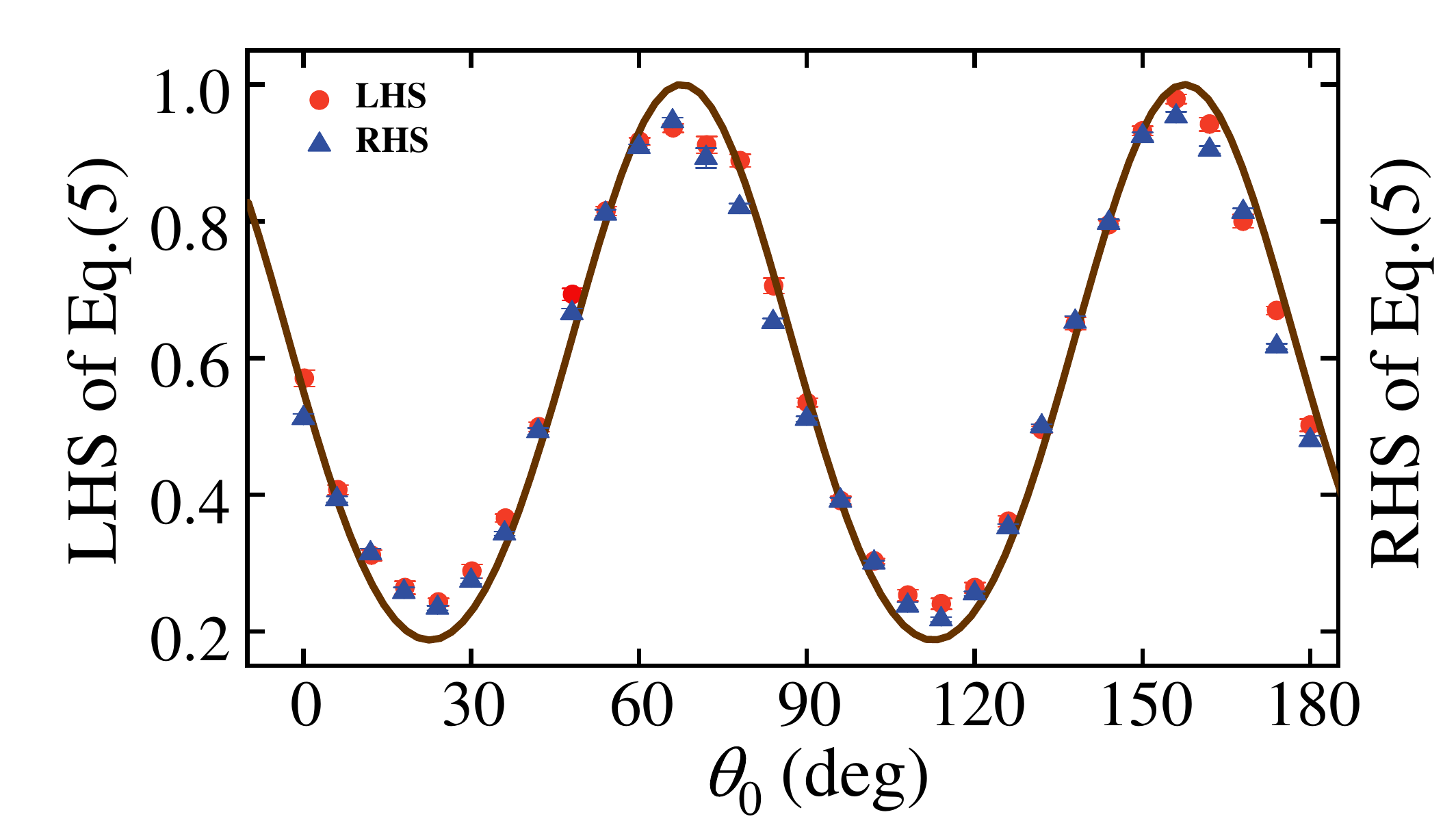}
\caption{Verification of the uncertainty relation for non-Hermitian operators in the form of Eq.~(\ref{UR6}), where both $A$ and $B$ are real. Using fixed non-Hermitian operators $A$ and $B$, we prepare different initial states. The left hand side (LHS) and right hand side (RHS) of Eq.~(\ref{UR6}) are red dots and blue triangles measured from experiments, respectively. The solid brown line represents the theoretical curve of Eq.~(\ref{UR6}).
}\label{Fig4}
\end{center}
\end{figure}

In the following, we denote $\theta_k$ as the angle of the half wave plate $\mathrm{H_k}$ (with $k$ from $0$ to $8$), and denote $\theta_{A}$ ($\theta_{B}$) as the  angle of the quarter wave plate $\mathrm{Q_A}$ ($\mathrm{Q_B}$). The half wave plate $\mathrm{H}_{0}$ allows a series of polarized qubit states to be encoded on the signal photon, which is the initial state
\begin{equation}
 |\varphi\rangle=\cos{2\theta_{0}}|0\rangle+\sin{2\theta_{0}|1\rangle},
\end{equation}
to be sent into the interference ring. We show two cases where the non-Hermitian operators (both $A$ and $B$) are (a) real or (b) complex. The matrix of a real non-Hermitian operator is given by all the real numbers in the chosen basis, whereas the matrix of a complex non-Hermitian operator consists  of complex numbers. It is worth noticing that we have polar decomposition  of the non-Hermitian operators $A=S_{A}U_{A}$ and $B=S_{B}U_{B}$ (we will use $A$ as an example, the same method applies to $B$). (a) In the real case, the unitary operator $U_{A}$ is realized by using the half wave plate $\mathrm{H}_{1}$, and the positive semidefinite operator $S_{A}$ is realized by using two beam displacement crystals $\mathrm{BD}_{1}$ and $\mathrm{BD}_{2}$ (the BD we selected is one step moved with horizontal polarization and the vertical polarization path remains unchanged) and half wave plates $\mathrm{H}_{2}$, $\mathrm{H}_{3}$ and $\mathrm{H}_{4}$. In this case,  we use the polar decomposition $A=S_{A}U_{A}$, with $S_{A}=\mathrm{diag}(-\cos{2\theta_{3}},1)$ and $U_{A}=\cos{\Delta\theta_{1,0}}|0\rangle\langle0|+\sin{\Delta\theta_{1,0}}|0\rangle\langle1|+\sin{\Delta\theta_{1,0}}|1\rangle\langle0|-\cos{\Delta\theta_{1,0}}|1\rangle\langle1|$, where we have defined $\Delta\theta_{1,0}:=2(\theta_{1}-\theta_{0})$. (b) In the complex case, the unitary operator $U_{A}$ is realized by using half wave plate $\mathrm{H}_{1}$ and quarter wave plate $\mathrm{Q_{A}}$, and by using $\mathrm{BD}_{1,2}$ and half wave plates $\mathrm{H}_{2}$, $\mathrm{H}_{3}$ and $\mathrm{H}_{4}$ we implement the positive semidefinite operator $S_{A}$, where $S_{A}=\mathrm{diag}(-\cos{2\theta_{3}},1)$ and $U_{A}=\cos{\Delta\theta_{1,0}}|0\rangle\langle0|+i\sin{\Delta\theta_{1,0}}|0\rangle\langle1|+\sin{\Delta\theta_{1,0}}|1\rangle\langle0|-i\cos{\Delta\theta_{1,0}}|1\rangle\langle1|$.
The treatment of the non-Hermitian operator $B$ is similar, with $U_{B}$ in the real case implemented by the half wave plate $\mathrm{H}_{5}$ and $S_{B}$ by $\mathrm{BD}_{3,4}$ as well as half wave plates $\mathrm{H}_{6}$, $\mathrm{H}_{7}$ and $\mathrm{H}_{8}$. The complex case $U_{B}$ is realized by the half wave plate $\mathrm{H}_{5}$ and the quarter wave plate $\mathrm{Q_{B}}$. $S_{B}$ is realized by BD crystals $\mathrm{BD}_{3}$ and $\mathrm{BD}_{4}$, and the half-wave plates $\mathrm{H}_{6}$, $\mathrm{H}_{7}$ and $\mathrm{H}_{8}$ (see Supplemental Material \cite{Supp}).

\textit{Results.---} The interference fringes in Fig.~\ref{Fig2} are obtained by measuring the photon counts at the output of the interferometer. Rotating the angle of the one-armed waveplate group in the interference ring gives the identity operator $I$, the non-Hermitian operator $A$, and the non-Hermitian operator $B$. Here we have chosen $I$ and $I$, and $B$ ($\theta_{5}=0^\circ$) and $I$, as examples to obtain the corresponding interference pattern at a given initial state $|\varphi\rangle$ ($\theta_{0}=-45^\circ$). The theoretical value of the interference visibility can reach $1$. In the experiment, the Sagnac interferometer visibility reached $98.28\%$. The $x$-axis represents the $\theta$ of rotation of the PS (once $0.04^\circ$), and the $y$-axis represents the number of recombination (per $1.5s$). The two curves (a) and (b) reflect the interference visibility of the interference loop in the experiment at a given initial state.

\begin{figure}
\begin{center}
\includegraphics[scale=0.20]{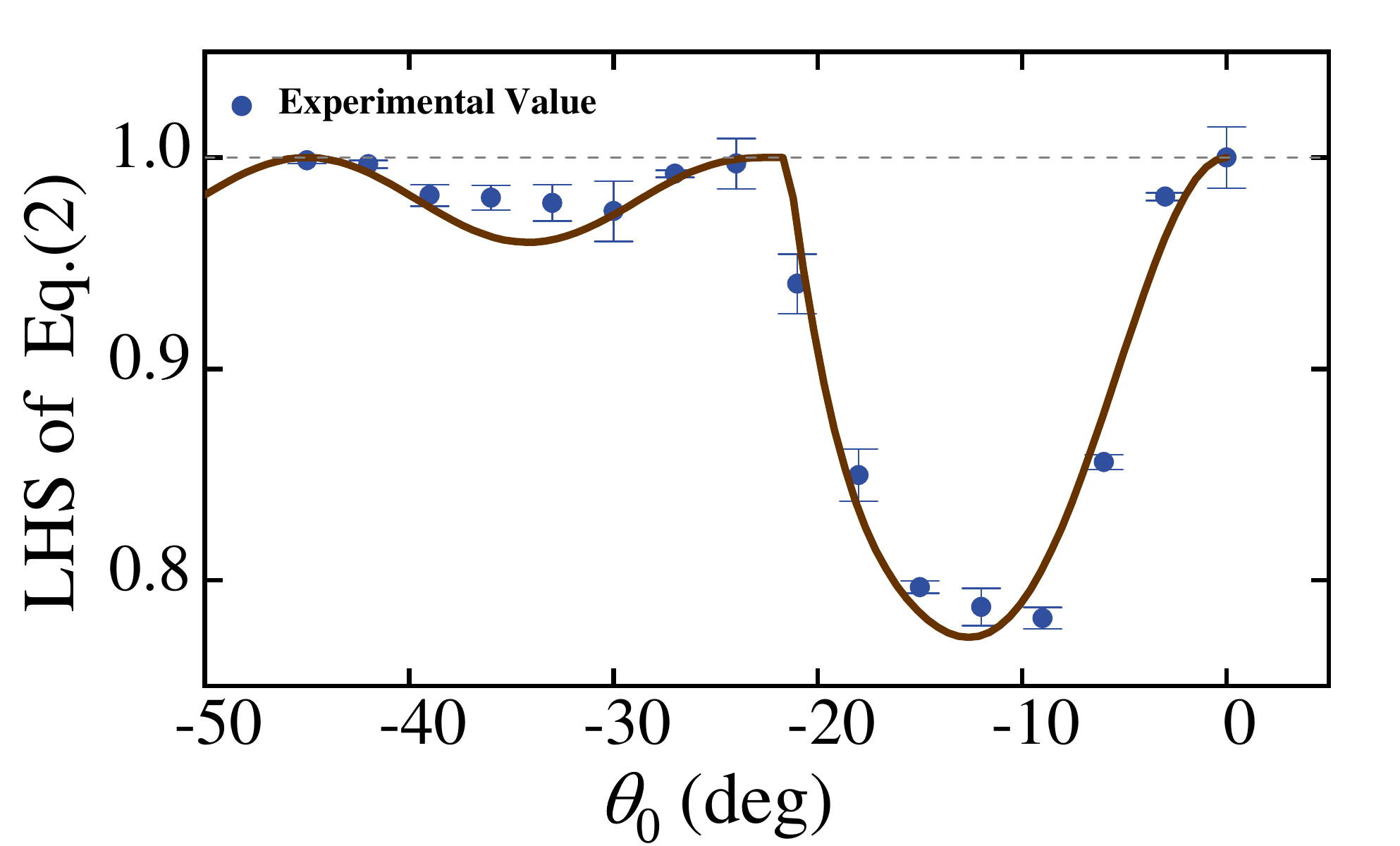}
\caption{Verification of the uncertainty relation for non-Hermitian operators in the form of Eq.~(\ref{UR5}). Ensure that non-Hermitian operators on both arms of Sagnac ring interferometer are fixed and input different initial states. The blue dot represents the experimental value of LHS in Eq.~(\ref{UR5}), and the brown solid line represents the theocratical value of LHS in Eq.~(\ref{UR5}). The maximum uncertainty state corresponds to a unit value indicated by a gray dashed line.}\label{Fig3}
\end{center}
\end{figure}

Testing the uncertainty relation of real non-Hermitian operators has been shown in Fig.~\ref{Fig1} with the real case (a). For real operators under a pure single-qubit state, the equality in Eq.~(\ref{UR5}) holds. We have
\begin{eqnarray}
\sum_{\substack{i,j,k=1\\i\neq j\neq k}}^3|T_{ij}|^2|T_{kk}|-2|T_{23} T_{12} T_{31}|=|T_{22} T_{33}|,\label{UR6}
\end{eqnarray}
where we have used $|T_{11}|=1$ and $|T_{ij}|=|T_{ji}|$. For the derivation of Eq.~(\ref{UR6}), please see Supplemental Material \cite{Supp}. Thus, in order to test Eq. (\ref{UR6}), we still only to measure $|T_{22}|$, $|T_{33}|$, $|T_{23}|$, $|T_{12}|$ and $|T_{31}|$.
For the construction of the non-Hermitian operators $A$ and $B$, we choose the following angles $\theta_{1}=22.5^\circ$, $\theta_{3}=60^\circ$, $\theta_{5}=22.5^\circ$, $\theta_{7}=75^\circ$ and $\theta_{2,4,6,8}=45^\circ$. The experimental results of Eq.~(\ref{UR6}) are shown in Fig.~\ref{Fig4}, where the $x$-axis represents the angle $\theta_{0}$ required for the preparation of the initial state $|\varphi\rangle$, and the $y$-axis represents the theoretical value range of Eq.~(\ref{UR6}). The theoretical curve is marked in brown and the experimental data points on the LHS and RHS of Eq.~(\ref{UR6}) are marked in red and blue, respectively. Due to the polar decomposition of non-Hermitian operators, the unitary operator in the case of real operators is relatively simple, and the results can be found in Supplemental Material \cite{Supp}. For non-Hermitian operators $A$ and $B$, details are derived in the supplementary material, giving the theoretical $|T_{ij}|$ ($i,j=1,2,3$) curve and the corresponding experimental results \cite{Supp}. According to the theoretical results, the left and right hand sides of Eq.~(\ref{UR6}) should be strictly equal. However, in our experiment there is a slight difference between the left and right hand sides, and one of the sources of error is the imperfect calibration of the waveplate, which causes the input states and non-Hermitian operators to deviate slightly from the expected setting.

The verification of the complex case of the non-Hermitian operator uncertainty relation in Eq.~(\ref{UR5}) is shown in Fig.~\ref{Fig3}. Here, we use the complex case (b) in Fig.~\ref{Fig1}, and let $\theta_{A,B}=0^\circ$, $\theta_{2,4,6,8}=45^\circ$ with fixed $A$ ($\theta_{1}=22.5^\circ$,$\theta_{3}=60^\circ$) and $B$ ($\theta_{5}=0^\circ$,$\theta_{7}=75^\circ$). Since QWP is added to construct the non-Hermitian operator of complex case, the unitary operator is slightly different from the real case. For details, please refer to Supplemental Material \cite{Supp}. The uncertainty relation can be tested from the measured values $|T_{ij}|$ in Eq.~(\ref{UR5}) for a pure input state. In practice, there are still some small experimental defects.
The experimental results have been shown as the $y$-axis, and the angles $\theta_0$ corresponding to initial states $|\varphi\rangle$ vary along the $x$-axis.
We use brown curve to mark the theoretical result and blue dots to represent the experimental data points. The terms in the inequality deviate slightly from the theoretical values in our experiment. This error results from the imperfect calibration of the wave plate and the initial state of the pure state. These imperfections cause the operator operation to differ from the expected setting. In the Supplemental Material, we provide the details for the case of a complex non-Hermitian operator.

The experimental results agree with the theoretical values. The possible sources of error include the imperfections of the prepared pure state, the imperfect separation of the horizontal and vertical polarizations by the BDs, and the deviation of the experimentally prepared non-Hermitian operator from the theoretical one.

\textit{Discussions and conclusions.---}
Our experiment is applicable in a number of areas.~Firstly, the uncertainty relations for non-Hermitian operators have been used to derive quantum Cram\'{e}r-Rao bound for arbitrary parameter-independent non-Hermitian Hamiltonians \cite{Xing}. 
Moreover, one can measure the fidelity of quantum states under quantum channels by non-Hermitian uncertainty relations in any dimension \cite{Pati}. 
Last but not least, the uncertainty relations for non-Hermitian operators can also be applied to entanglement detection (see Sec. IV for details in \cite{Supp}), which is generalized from Ref. \cite{Hofm}. All these applications of uncertainty relations for non-Hermitian operators can be realized in experiments based on our methods.

Our method can be connected with the non-Hermitian operators $H_{PT}$ based on metric operators. 
One can also use our experimental method to mesure $|\langle H_{PT}\rangle|$, where $H_{PT}$ satisfys $H_{PT}^\dag=\eta H_{PT}\eta^{-1}$ with a proper metric operator $\eta$ \cite{Most,Nori,Sand}. Compared with the decomposition method $h_{PT}=\eta^{1/2}H_{PT}\eta^{-1/2}$ where $h_{PT}$ is Hermitian \cite{Sand}, it is more convenient to realize $H_{PT}$ operation in linear optical experiments by using our method.

We thank Arun Kumar Pati and Xiao-Min Hu for discussions. This work is supported by the Innovation Program for Quantum Science and Technology (Grants No. 2021ZD0301200, No.2021ZD0301400), the National Natural Science Foundation of China (Grants No.~11734015, No.~11821404 and No.~11875172), and K.C. Wong Magna Fund in Ningbo University.

\setcounter{equation}{0}
\renewcommand{\theequation}{S\arabic{equation}}
\onecolumngrid

\section*{Appendix A. Theoretical methods}

\subsection{Theory of uncertainty relations for non-Hermitian operators}
The uncertainty relation of two non-Hermitian operators is expressed as
\begin{eqnarray}
\langle(\Delta A)^2\rangle \langle (\Delta B)^2\rangle \geq |\langle A^\dag B\rangle-\langle A^\dag\rangle \langle B\rangle|^2.\label{U1}
\end{eqnarray}
where $\langle(\Delta A)^2\rangle:=\langle A^\dag A\rangle-\langle A^\dag \rangle \langle A \rangle$, and $\langle(\Delta B)^2\rangle:=\langle B^\dag B\rangle-\langle B^\dag \rangle \langle B \rangle$ \cite{Pati,Zhao,Anna}.

The uncertainty relation of two non-Hermitian operators in a $d$-dimensional Hilbert space is shown in Eq.~(\ref{U1}), we consider the Eq.~(\ref{U1}) in pure states, we define the state is a pure state $|\psi\rangle$, then
\begin{eqnarray}
\langle(\Delta A)^2\rangle=\langle\psi|A^\dag A|\psi\rangle-\langle\psi|A^\dag|\psi\rangle\langle\psi|A|\psi\rangle=\langle\psi|A^\dag PA|\psi\rangle,\label{U11}
\end{eqnarray}
where $P$ is a project operator defined as $P:=1-|\psi\rangle\langle\psi|$. Similarly, the variance of $B$ can be written as
\begin{eqnarray}
\langle(\Delta B)^2\rangle=\langle\psi|B^\dag B|\psi\rangle-\langle\psi|B^\dag|\psi\rangle\langle\psi|B|\psi\rangle=\langle\psi|B^\dag PB|\psi\rangle,\label{U12}
\end{eqnarray}
and
\begin{eqnarray}
|\langle A^\dag B\rangle-\langle A^\dag\rangle\langle B\rangle|^2=|\langle\psi|A^\dag B|\psi\rangle-\langle\psi|A^\dag|\psi\rangle\langle\psi|B|\psi\rangle|^2=|\langle\psi|A^\dag PB|\psi\rangle|^2.\label{U13}
\end{eqnarray}

Hence, the Eq.~(\ref{U1}) with a pure state $|\psi\rangle$ become
\begin{eqnarray}
\langle\psi|A^\dag PA|\psi\rangle\langle\psi|B^\dag PB|\psi\rangle\geq|\langle\psi|A^\dag PB|\psi\rangle|^2.\label{U14}
\end{eqnarray}

Furthermore, if $d=2$ (for a single-qubit system), the rank of $P$ is 1. Thus, $P$ can be rewritten as $P=|\psi_{\bot}\rangle\langle\psi_{\bot}|$, where $|\psi_{\bot}\rangle$ is the orthogonal state of $|\psi\rangle$ in the single qubit system. So the Eq.~(\ref{U14}) becomes
\begin{eqnarray}
\langle\psi|A^\dag|\psi_{\bot}\rangle\langle\psi_{\bot} |A|\psi\rangle\langle\psi|B^\dag|\psi_{\bot}\rangle\langle\psi_{\bot}|B|\psi\rangle=|\langle\psi|A^\dag|\psi_{\bot}\rangle|^2|\langle\psi|B|\psi_{\bot}\rangle|^2=|\langle\psi|A^\dag|\psi_{\bot}\rangle\langle\psi_{\bot}|B|\psi\rangle|^2.\label{U15}
\end{eqnarray}

Therefore, it is clear that the $``="$ always hold in Eqs.~(\ref{U14}) and (\ref{U1})  when we consider a single qubit pure state.

\subsection{Derivation of Eq.~(2) in the main text}
Now we derive Eq.~(2) in the main text.
We can rewrite Eq.~(\ref{U1}) as
\begin{eqnarray}
\langle B^\dag B\rangle\langle A^\dag\rangle\langle A\rangle+\langle A^\dag A\rangle\langle B^\dag\rangle\langle B \rangle+\langle A^\dag B\rangle\langle B^\dag A\rangle-\langle A^\dag B\rangle\langle A\rangle\langle B^\dag\rangle-\langle B^\dag A\rangle\langle B\rangle\langle A^\dag\rangle\leq \langle A^\dag A\rangle \langle B^\dag B\rangle,\label{U2}
\end{eqnarray}
if $\langle A^\dag A\rangle \langle B^\dag B\rangle\neq0$ we have
\begin{eqnarray}
\cfrac{\langle A^\dag\rangle\langle A \rangle}{\langle A^\dag A\rangle}+\cfrac{\langle B^\dag\rangle\langle B \rangle}{\langle B^\dag B\rangle}+\cfrac{\langle A^\dag B\rangle\langle B^\dag A\rangle}{\langle A^\dag A\rangle\langle B^\dag B\rangle}-\cfrac{\langle A^\dag B\rangle\langle A\rangle\langle B^\dag\rangle}{\langle A^\dag A\rangle\langle B^\dag B\rangle}-\cfrac{\langle B^\dag A\rangle\langle B\rangle\langle A^\dag\rangle}{\langle A^\dag A\rangle\langle B^\dag B\rangle}\leq 1.\label{U3}
\end{eqnarray}
Let us define $|\phi_{1}\rangle :=|\varphi\rangle$, $|\phi_{2}\rangle :=A|\varphi\rangle$, $|\phi_{3}\rangle:=B|\varphi\rangle$, and $T_{ij}:=\langle\phi_{i}|\phi_{j}\rangle$ with $(i,j=1,2,3)$, the uncertainty relation Eq.~(\ref{U2}) becomes
\begin{eqnarray}
|T_{12}|^2|T_{33}|+|T_{13}|^2|T_{22}|+|T_{23}|^2-T_{23}T_{12}T_{31}-T_{32}T_{21}T_{13}\leq |T_{22} T_{33}|.\label{U4}
\end{eqnarray}
Since $T_{32}T_{21}T_{13}=(T_{23}T_{12}T_{31})^*$, we have
\begin{eqnarray}
T_{23}T_{12}T_{31}&:=& |T_{23}T_{12}T_{31}|e^{i\Phi},\\
\mathrm{Re}(T_{23}T_{12}T_{31})&=&\frac{T_{23}T_{12}T_{31}+T_{32}T_{21}T_{13}}{2}=|T_{23}T_{12}T_{31}|\cos\Phi,
\end{eqnarray}
where $\Phi$ is the phase of $T_{23}T_{12}T_{31}$, and $\mathrm{Re}(T_{23}T_{12}T_{31})$ is the real part of $T_{23}T_{12}T_{31}$. Thus, the Eq.~(\ref{U4}) is equivalent to
\begin{eqnarray}
|T_{12}|^2|T_{33}|+|T_{13}|^2|T_{22}|+|T_{23}|^2-|T_{22}T_{33}|&\leq& 2|T_{23}T_{12}T_{31}|\cos\Phi.\label{U5}
\end{eqnarray}
From $\cos\Phi\leq1$, one has
\begin{eqnarray}
|T_{12}|^2|T_{33}|+|T_{13}|^2|T_{22}|+|T_{23}|^2-|T_{22}T_{33}|&\leq& 2|T_{23}T_{12}T_{31}|.\label{U52}
\end{eqnarray}
If $|T_{22}T_{33}|\neq0$, we have a weaker uncertainty relation of ~(\ref{U5}),
\begin{eqnarray}
\frac{|T_{12}|^2}{|T_{22}|}+\frac{|T_{13}|^2}{|T_{33}|}+\frac{|T_{23}|^2}{|T_{22}T_{33}|}-2\frac{|T_{23}T_{12}T_{31}|}{|T_{22}T_{33}|}\leq 1.\label{U6}
\end{eqnarray}

\subsection{Derivation of Eq. (5) in the main text}
Now we derive Eq. (5) in the main text. On the above, we have derived that Eq. (\ref{U1}) is equivalent to Eq. (\ref{U4}). If $2|T_{23}T_{12}T_{31}|=T_{23}T_{12}T_{31}+T_{32}T_{21}T_{13}$ holds, Eq. (\ref{U1}) is equivalent to
\begin{equation}
    |T_{12}|^2|T_{33}|+|T_{13}|^2|T_{22}|+|T_{23}|^2-2|T_{23}T_{12}T_{31}|\leq |T_{22} T_{33}|.\label{UU1}
\end{equation}
Now we discuss when $``="$ in Eq. (\ref{UU1}) holds. The first condition is that $2|T_{23}T_{12}T_{31}|=T_{23}T_{12}T_{31}+T_{32}T_{21}T_{13}$ should hold, and thus Eq. (\ref{UU1}) is equivalent to Eq. (\ref{U1}). Meanwhile, the second condition is that  $``="$ in Eq. (\ref{U1}) should hold, i.e., we should consider the single qubit pure states, as discussed in Eq. (\ref{U15}). Therefore, if we consider a single qubit pure state $|\psi\rangle$ such that  $2|T_{23}T_{12}T_{31}|=T_{23}T_{12}T_{31}+T_{32}T_{21}T_{13}$ holds, Eq. (\ref{U1}) is equivalent to Eq.(5) in the main text, i.e.,
\begin{equation}
   \sum_{\substack{i,j,k=1\\i\neq j\neq k}}^3|T_{ij}|^2|T_{kk}|-2|T_{23} T_{12} T_{31}|= |T_{12}|^2|T_{33}|+|T_{13}|^2|T_{22}|+|T_{23}|^2-2|T_{23}T_{12}T_{31}|= |T_{22} T_{33}|.\label{UU2}
\end{equation}

In the experiment, we verify the cases of real and complex numbers, respectively. In the case of real numbers, we have chosen a special scenario, where the single-qubit pure states and operators are all real such that  $2|T_{23}T_{12}T_{31}|=T_{23}T_{12}T_{31}+T_{32}T_{21}T_{13}$ holds. On this basis, we verify that Eq.~(\ref{UU2}) make sense.

In the complex case, as $\mathrm{Re}(T_{23}T_{12}T_{31})<|T_{23}T_{12}T_{31}|$, we experimentally verify the weaker version of the uncertainty relation, i.e.,  Eq.~(\ref{U6}).

\subsection{Experiment of uncertainty relations for non-Hermitian operators}
\begin{figure*}
\includegraphics[width=10cm]{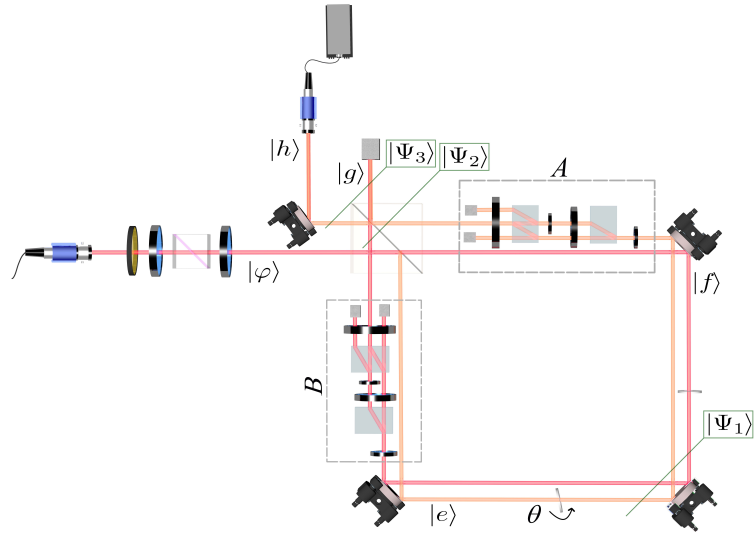}
\caption{An experimental setup for testing the uncertainty relations between two non-Hermitian operators $A$ and $B$ by using the Sagnac ring interferometer. We use a $50:50$ non-polarizing beam splitter (BS) to construct a shifted Sagnac ring interferometer, the non-Hermite operators $A$ and $B$ act on the reflection (orange) and transmission (red) paths respectively.} \label{FigX}
\end{figure*}

Because some photons are lost, but normalisation must be ensured, we use analogues that interfere with visibility. We defined that $|\varphi\rangle=\cos{2\theta_{0}}|0\rangle+\sin{2\theta_{0}|1\rangle}$ is the initial state before the BS, with $|0\rangle:=|H\rangle$ being the horizontal polarization state  and $|1\rangle:=|V\rangle$ being the vertical polarization state. The state after the BS changes into
\begin{eqnarray}
|\Psi_{1}\rangle=\dfrac{1}{\sqrt{2}}\Big(ie^{i\theta}|e\rangle+|f\rangle\Big)|\varphi\rangle=\dfrac{1}{\sqrt{2}}\Big(ie^{i\theta}|e\rangle+|f\rangle\Big)\Big(\cos{2\theta_{0}}|0\rangle+\sin{2\theta_{0}}|1\rangle\Big),\label{UU6}
\end{eqnarray}
where $|e\rangle$ and $|f\rangle$ denote the reflection and transmission paths of the Sagnac interferometer respectively. As shown in Fig.~\ref{FigX}, we operated on both arms with a series of optical elements for the non-Hermitian operators $A$ and $B$ (grey dashed boxes), Eq.~(\ref{U6}) is further change as
\begin{eqnarray}
|\Psi_{2}\rangle=\dfrac{1}{\sqrt{2}}\Big(ie^{i\theta}A|\varphi\rangle|e\rangle+B|\varphi\rangle|f\rangle\Big).\label{U7}
\end{eqnarray}

After passing through the BS again, the two paths are called $|g\rangle$ and $|h\rangle$, we got
\begin{eqnarray}
|\Psi_{3}\rangle=\dfrac{1}{2}\Big(-e^{i\theta}A|\varphi\rangle+B|\varphi\rangle\Big)|g\rangle+\dfrac{1}{2}\Big(ie^{i\theta}A|\varphi\rangle+iB|\varphi\rangle\Big)|h\rangle.\label{U8}
\end{eqnarray}

The SPD is at port $|h\rangle$, which means that detected a fraction of the total state. This can be carried out by applying the projector $\Pi_{h}=|h\rangle\langle h|$ to entire state, so the component of the detection arm state is
\begin{eqnarray}
|\Psi_{h}\rangle=\Pi_{h}|\Psi_{3}\rangle=\dfrac{1}{2}\Big(ie^{i\theta}A|\varphi\rangle+iB|\varphi\rangle\Big)|h\rangle.\label{U9}
\end{eqnarray}
Finally, the strength of the detector port is determined by
\begin{eqnarray}
N_{h}=|\langle h|\Psi_{h}\rangle|^2=\dfrac{1}{4}\Big(\langle A^\dag A\rangle+\langle B^\dag B\rangle+e^{-i\theta}\langle A^\dag B\rangle+e^{i\theta}\langle B^\dag A\rangle\Big)=\dfrac{1}{4}\Big(\langle A^\dag A\rangle+\langle B^\dag B\rangle+2|\langle A^\dag B\rangle|\cos(\psi-\theta)\Big),\label{U10}
\end{eqnarray}
where we suppose that $A^\dag B=|A^\dag B|e^{i\psi}$ and thus $B^\dag A=|A^\dag B|e^{-i\psi}$.

Hence, we can obtain the analogues that interfere with visibility $|T_{ij}|$. Since the final state must be normalized, the maximum and minimum values of $N_{h}$ must be divided by the total number of photons. Theoretically, the total number of photons should be the maximum value $I_{max}$ of the unit operator operated by both the reflection path and the transmission path (the theoretical value of $I_{min}=0$). However, in the actual experiment $I_{min}\neq0$, so the total number of photons should be $I_{max}+I_{min}$. Based on Eq.~(\ref{U10}), the maximum and minimum values of $N_{h}$ after normalization can be obtained by varying $\theta$,
\begin{eqnarray}
\dfrac{(N_{h})_{max}}{I_{max}+I_{min}}=\dfrac{1}{4}(\langle A^\dag A\rangle+\langle B^\dag B\rangle+2|\langle A^\dag B\rangle|),\\\label{U11}
\dfrac{(N_{h})_{min}}{I_{max}+I_{min}}=\dfrac{1}{4}(\langle A^\dag A\rangle+\langle B^\dag B\rangle-2|\langle A^\dag B\rangle|).\label{U12}
\end{eqnarray}

The analogues that interfere with visibility $|T_{ij}|$ is defined as
\begin{eqnarray}
|T_{ij}|:=\dfrac{(N_{h})_{max}-(N_{h})_{min}}{I_{max}+I_{min}}=|\langle A^\dag B\rangle|.\label{U13}
\end{eqnarray}

\begin{figure*}
\includegraphics[width=8cm]{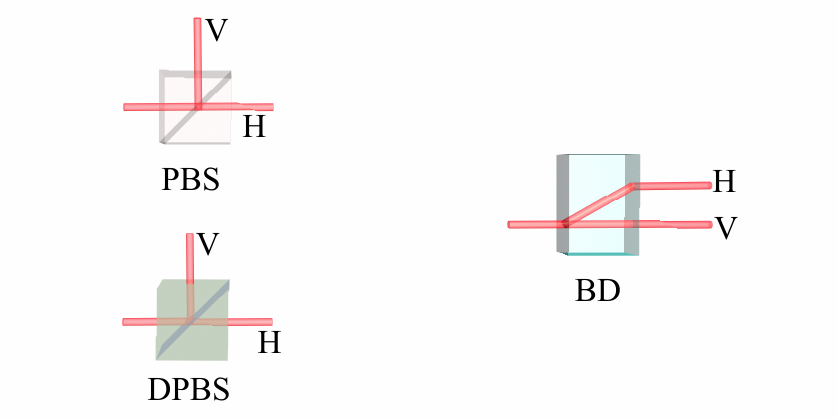}
\caption{An example of a crystal with path and polarization information.} \label{FigI}
\end{figure*}

\subsection{Error analysis}
In this study, we choose the error propagation formula to solve data from the experiment.
We can suppose that the function of the indirect measurement $y$ and the direct measurement $x_{1}, \cdots, x_{N}$ as \cite{Jhon}
\begin{equation}
y=f(x_{1}, \cdots, x_{N}), \label{F1}
\end{equation}

By expanding Eq.~(\ref{F1}) in $Taylor$ $series$ around the expected value $x_{1}, \cdots, x_{N}$ of $\mu_{1}, \cdots, \mu_{N}$, ignoring terms above second order, we have
\begin{equation}
y \approx f(\mu_{1}, \cdots, \mu_{N})+\sum\limits_{i=1}^N(\frac{\partial f}{\partial x_{i}})_{\mu_{1}\ldots\mu_{N}} (x_{i}-\mu_{i}). \label{F2}
\end{equation}
and take the expectation on both sides
\begin{equation}
E(y)\approx f(\mu_{1},\cdots \mu_{N}). \label{F3}
\end{equation}
We put Eq.~(\ref{F2}) in Eq.~(\ref{F3}), we have
\begin{equation}
[y-E(y)]^2 \approx \sum\limits_{i=1}^N(\frac{\partial f}{\partial x_{i}})_{\mu_{1}\ldots\mu_{N}}^2 (x_{i}-\mu_{i})^2+2\sum\limits_{i=1}^N\sum\limits_{j=i+1}^{N}(\frac{\partial f}{\partial x_{i}})_{\mu_{1}\ldots\mu_{N}}(\frac{\partial f}{\partial x_{j}})_{\mu_{1}\ldots\mu_{N}} (x_{i}-\mu_{i})(x_{j}-\mu_{j}). \label{F4}
\end{equation}

Take the expected value on both sides, and you get the variance transfer formula that we used in the experiment to deal with the data
\begin{equation}
\sigma^2(y)\approx\sum\limits_{i=1}^N(\frac{\partial f}{\partial x_{i}})_{\mu_{1}\ldots\mu_{N}}^2 \sigma^2(x_{i})+2\sum\limits_{i=1}^N\sum\limits_{j=i+1}^{N}(\frac{\partial f}{\partial x_{i}})_{\mu_{1}\ldots\mu_{N}} (\frac{\partial f}{\partial x_{j}})_{\mu_{1}\ldots\mu_{N}} Cov(x_{i},x_{j}). \label{F5}
\end{equation}

We take
\begin{equation}
|T_{23}|=\cfrac{N_{max}(A^\dag,B)-N_{min}(A^\dag,B)}{N_{max}(I,I)+N_{min}(I,I)}, 
\end{equation}as an example, the relation between the variables $x_{1}$ and $x_{2}$ can be written as $y=x_{1}/x_{2}$, $x_{1}:N_{max}(A^\dag,B)-N_{min}(A^\dag,B)$, $x_{2}:N_{max}(I,I)+N_{min}(I,I)$ and $y:|T_{23}|$. The Eq.~(\ref{F5}) can be written as
\begin{equation}
\sigma^2(y)=\Bigg(\frac{1}{\overline{x_{2}}}\Bigg)^2\sigma^2(x_{1})+\Bigg(\frac{\overline{x_{1}}}{\overline{x_{2}}^2}\Bigg)^2\sigma^2(x_{2})+2\Bigg(\frac{1}{\overline{x_{2}}}\Bigg)\Bigg(\frac{\overline{x_{1}}}{\overline{x_{2}}^2}\Bigg)Cov(x_{1},x_{2}). \label{F}
\end{equation}
Since the variables $x_{1}$ and $x_{2}$ are independent of each other, so $Cov(x_{1},x_{2})=0$. Note that the variance $\sigma^2(x_{i})$ should be calculated as the sum of the variances of the measured data in each variable $x_{i}$ ($i=1,2$). Because of the variable itself contains measurement errors and according to our theory, the function that includes relevant variables is also expected to have errors due to their influence. Therefore, it is essential to utilize the error propagation formula when handling the experimental data.

\section*{Appendix B. The scheme of the experiment}
Before detailing our experimental procedure, we first introduce the important components in the optical components:

$1).$ We use half-wave plates (HWP: $\alpha$) and quarter-wave plates (QWP: $\beta$) to implement non-Hermitian operation. The $\alpha$ or $\beta$ here refers to the angle between the fast axis of the waveplate and the horizontal polarization direction. The $Jones$ $matrix$ of waveplates can be denoted as \cite{Jame}
\begin{eqnarray}
HWP(\alpha)=
   \begin{bmatrix}
     \cos2\alpha&\sin2\alpha\\
     \sin2\alpha&-\cos2\alpha
   \end{bmatrix},
QWP(\beta)=
   \begin{bmatrix}
     \cos^2\beta+i\sin^2\beta&(1-i)\cos\beta\sin\beta\\
     (1-i)\cos\beta\sin\beta&\sin^2\beta+i\cos^2\beta
   \end{bmatrix}
\end{eqnarray}

$2).$ We have replaced the horizontal polarization state $|H\rangle$ and the vertical polarization state $|V\rangle$ as follows
\begin{eqnarray}
|H\rangle=|0\rangle=
   \begin{bmatrix}
    1\\
     0
   \end{bmatrix},
|V\rangle=|1\rangle=
   \begin{bmatrix}
    0\\
     1
   \end{bmatrix}
\end{eqnarray}

$3).$ The beam displacer (BD) is capable of fully transmitting horizontally polarized photons, but diverting them from their original path when they emit vertically polarized photons ($4mm$ in our experiments), see Fig.~\ref{FigI}.

$4).$ Polarization beam splitter (PBS) (dual-wave PBS (DPBS)) can divide polarized light into two paths, namely horizontal polarized light transmission and vertical polarized light reflection, see Fig.~\ref{FigI}.

\begin{figure*}
\includegraphics[width=9cm]{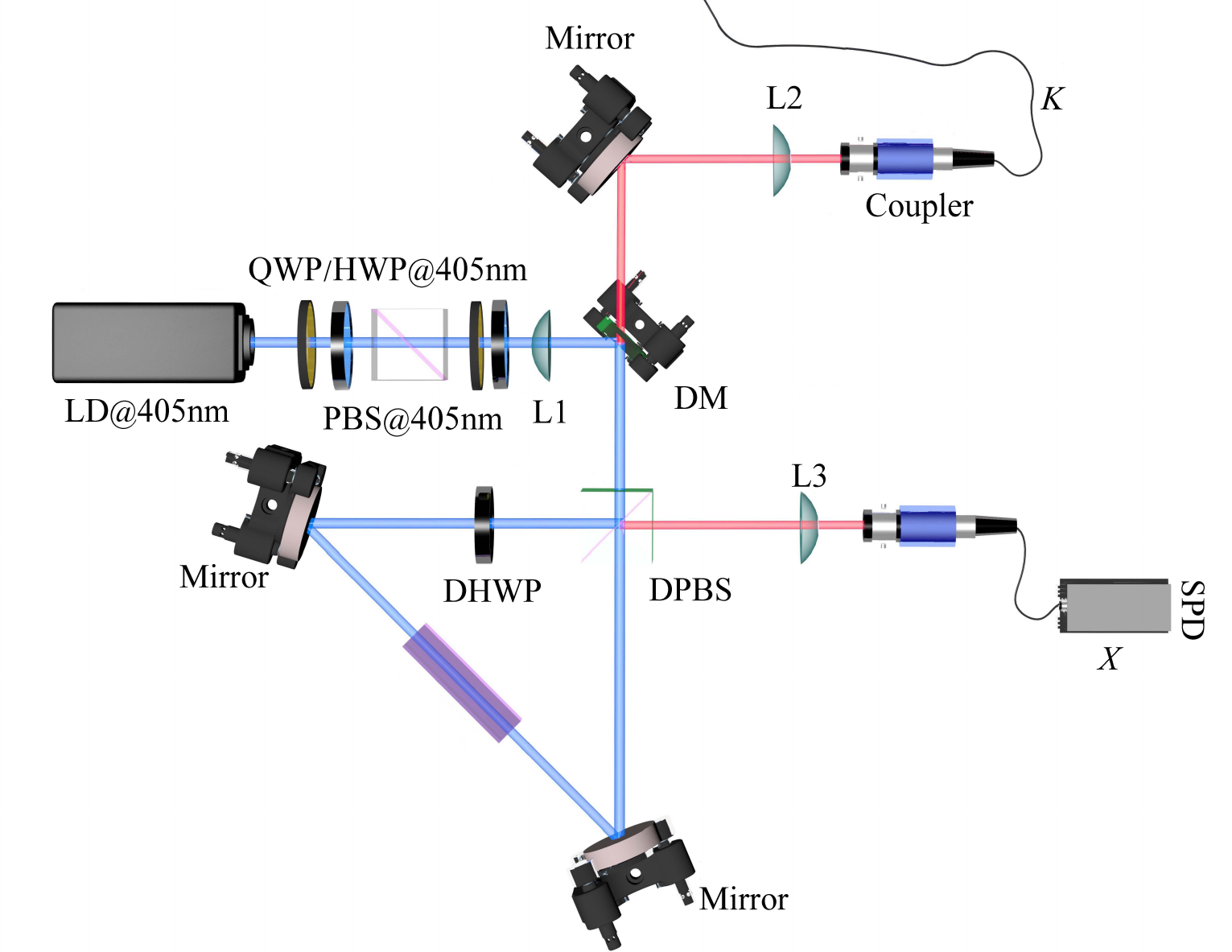}
\caption{Experimental setup for a heralded single-photon source of a Sagnac loop.} \label{FigII}
\end{figure*}

\subsection{The heralded single-photon source}
The pump light required for the experiment is generated by an ultraviolet (UV) laser diode with a central wavelength of $405$ nm. The power of the diode laser can be adjusted by the wave plate group (HWP and QWP) and PBS. Then HWP and QWP are used to prepare the polarization state of the diode laser. Here, HWP is rotated $0^\circ$ or $45^\circ$ (the angle between the fast axis and the horizontal direction). Due to the small size of the PPKTP crystal, two lenses $f_{1}=50$ mm and $f_{2}=100$ mm are used to focus the diode laser (as shown in Fig.~\ref{FigII} for L1, the L1 set of lenses is matched to a wavelength of $405$ nm). The dichroic polarization beam splitter (DPBS) can separate different polarization states and enter the PPKTP crystal simultaneously in both clockwise and counterclockwise directions \cite{Baos,Teah}. The working temperature of PPKTP crystal is set at $25^{\circ}C$, and the photon with wavelength of $810nm$ is generated by type-$II$ spontaneous down-conversion. The dichroic half wave plate (DHWP) is set in the counterclockwise direction and transforms $|V_{p}\rangle \rightarrow |H_{p}\rangle$. The effect of the dichroic mirror (DM) is that photons of $405$ nm are reflected and photons of $810$ nm are transmitted. The resulting photon pairs are collected into the single-mode fiber by optical collimator after passing through the lens set L2 and L3 (L2 and L3 set of lenses is matched to a wavelength of $810nm$). The $K$ in Fig.~\ref{FigII} represents the signal photon entering the actual optical path, which is finally counted by SPD(Y) and SPD(X). Since only a single photon source is needed in this experiment, we do not make any special requirements for the biased entanglement of this light source.

\subsection{Preparation of non-Hermitian operators A and B}
The primed photons are injected into an operator preparation module, consisting of several waveplates and two beam displacers (BDs), to produce the non-Hermitian operators required for our experiments. The size of the BDs allows two polarisations to be separated by 4 mm.
\begin{figure*}[b]
\includegraphics[width=12cm]{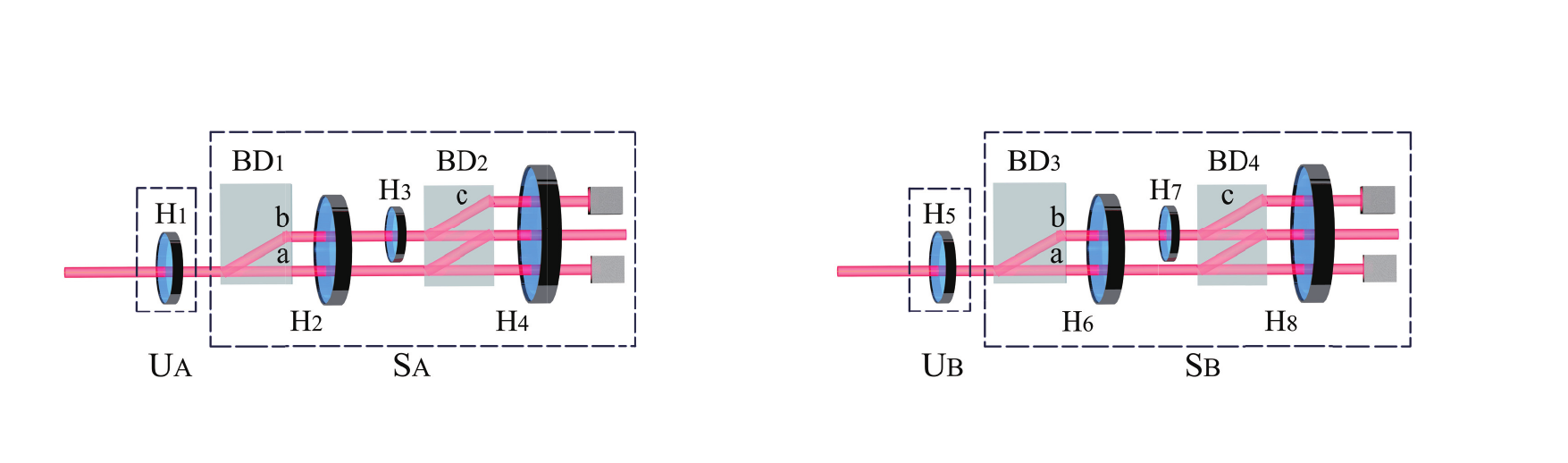}
\caption{Experimental setup of non-Hermitian operators $A=S_{A}U_{A}$ and $B=S_{B}U_{B}$ in the case of real numbers.}\label{FigIII}
\end{figure*}

The two cases where the non-Hermitian operator is real and the non-Hermitian operator is complex were chosen in our experiments, the non-Hermitian operator $A=S_{A}U_{A}$, $B=S_{B}U_{B}$. As shown in Fig.~\ref{FigIII}, we first present the preparation of the non-Hermitian operator $A$ in the real case. The process implementation can be shown in Table~\ref{tab1}.
\begin{table*}[htbp]
	\centering
	\caption{The preparation of the non-Hermitian operator $A$ in the real case.}
\resizebox{\linewidth}{!}{
	\begin{tabular}{cccc}
     \hline
     \hline
		State prepared & \ \ \ \ \ \ \ \ \ \ \ \ \ \ \ \ \ \ \ \ \ \ \ \ \ \ \ \  & Result  \\	
  \hline
$H_{0}$ & \ \ \ \ \ \ \ \ \ \ \ \ \ \ \ \ \ \ \ \ \ \ \ \ \ \ \ \  & $\cos2\theta_{0}|0\rangle+\sin2\theta_{0}|1\rangle$ \\
$H_{1}$ & \ \ \ \ \ \ \ \ \ \ \ \ \ \ \ \ \ \ \ \ \ \ \ \ \ \ \ \  & $\cos2(\theta_{1}-\theta_{0})|0\rangle+\sin2(\theta_{1}-\theta_{0})|1\rangle$\\
$BD_{1}$ & \ \ \ \ \ \ \ \ \ \ \ \ \ \ \ \ \ \ \ \ \ \ \ \ \ \ \ \  & $\cos2(\theta_{1}-\theta_{0})|0\rangle|b\rangle+\sin2(\theta_{1}-\theta_{0})|1\rangle|a\rangle$\\
$H_{2}$ ($45^\circ$) & \ \ \ \ \ \ \ \ \ \ \ \ \ \ \ \ \ \ \ \ \ \ \ \ \ \ \ \  & $\cos2(\theta_{1}-\theta_{0})|1\rangle|b\rangle+ \sin2(\theta_{1}-\theta_{0})|0\rangle|a\rangle$\\
$H_{3}$ & \ \ \ \ \ \ \ \ \ \ \ \ \ \ \ \ \ \ \ \ \ \ \ \ \ \ \ \  & $\sin2\theta_{3}\cos2(\theta_{1}-\theta_{0})|0\rangle|b\rangle-\cos2\theta_{3}\cos2(\theta_{1}-\theta_{0})|1\rangle|b\rangle+\sin2(\theta_{1}-\theta_{0})|0\rangle|a\rangle$\\
$BD_{2}$ & \ \ \ \ \ \ \ \ \ \ \ \ \ \ \ \ \ \ \ \ \ \ \ \ \ \ \ \  & $(-\cos2\theta_{3}\cos2(\theta_{1}-\theta_{0})|1\rangle+\sin2(\theta_{1}-\theta_{0})|0\rangle)|b\rangle+\sin2\theta_{3}\cos2(\theta_{1}-\theta_{0})|0\rangle|c\rangle$\\
$H_{4}$ ($45^\circ$) & \ \ \ \ \ \ \ \ \ \ \ \ \ \ \ \ \ \ \ \ \ \ \ \ \ \ \ \ & $-\cos2\theta_{3}\cos2(\theta_{1}-\theta_{0})|0\rangle|b\rangle+\sin2(\theta_{1}-\theta_{0})|1\rangle|b\rangle$\\
     \hline
     \hline
	\end{tabular}
}\label{tab1}
\end{table*}
Since we have done a polarization decomposition of the non-Hermitian operator $A$, we got $-\cos2\theta_{3}\cos2(\theta_{1}-\theta_{0})|0\rangle+\sin2(\theta_{1}-\theta_{0})|1\rangle$, this corresponds to $U_{A}$ and $S_{A}$ in the following matrix form
\begin{eqnarray}
U_{A}=
   \begin{bmatrix}
     \cos2(\theta_{1}-\theta_{0})&\sin2(\theta_{1}-\theta_{0})\\
     \sin2(\theta_{1}-\theta_{0})&-\cos2(\theta_{1}-\theta_{0})
   \end{bmatrix},
S_{A}=
   \begin{bmatrix}
     -\cos2\theta_{3}&0\\
     0&1
   \end{bmatrix}.\label{M1}
\end{eqnarray}


Similar to the operation of the non-Hermitian operator $A$, the right-hand side of Fig.~\ref{FigIII} shows the corresponding experimental design for the non-Hermitian operator $B$. The details of the process implementation can be found in Table~\ref{tab2}.
\begin{table*}[htbp]
	\centering
	\caption{The preparation of the non-Hermitian operator $B$ in the real case.}
\resizebox{\linewidth}{!}{
	\begin{tabular}{cccc ccc}
     \hline
     \hline
		State prepared & \ \ \ \ \ \ \ \ \ \ \ \ \ \ \ \ \ \ \ \ \ \ \ \ \ \ \ \  & Result  \\	
  \hline
$H_{0}$ & \ \ \ \ \ \ \ \ \ \ \ \ \ \ \ \ \ \ \ \ \ \ \ \ \ \ \ \  & $\cos2\theta_{0}|0\rangle+\sin2\theta_{0}|1\rangle$ \\
$H_{5}$ & \ \ \ \ \ \ \ \ \ \ \ \ \ \ \ \ \ \ \ \ \ \ \ \ \ \ \ \  & $\cos2(\theta_{5}-\theta_{0})|0\rangle+\sin2(\theta_{5}-\theta_{0})|1\rangle$\\
$BD_{3}$ & \ \ \ \ \ \ \ \ \ \ \ \ \ \ \ \ \ \ \ \ \ \ \ \ \ \ \ \  & $\cos2(\theta_{5}-\theta_{0})|0\rangle|b\rangle+\sin2(\theta_{5}-\theta_{0})|1\rangle|a\rangle$\\
$H_{6}$ ($45^\circ$) & \ \ \ \ \ \ \ \ \ \ \ \ \ \ \ \ \ \ \ \ \ \ \ \ \ \ \ \  & $\cos2(\theta_{5}-\theta_{0})|1\rangle|b\rangle+ \sin2(\theta_{5}-\theta_{0})|0\rangle|a\rangle$\\
$H_{7}$ & \ \ \ \ \ \ \ \ \ \ \ \ \ \ \ \ \ \ \ \ \ \ \ \ \ \ \ \  & $\sin2\theta_{7}\cos2(\theta_{5}-\theta_{0})|0\rangle|b\rangle-\cos2\theta_{7}\cos2(\theta_{5}-\theta_{0})|1\rangle|b\rangle+\sin2(\theta_{5}-\theta_{0})|0\rangle|a\rangle$\\
$BD_{4}$ & \ \ \ \ \ \ \ \ \ \ \ \ \ \ \ \ \ \ \ \ \ \ \ \ \ \ \ \  & $(-\cos2\theta_{7}\cos2(\theta_{5}-\theta_{0})|1\rangle+\sin2(\theta_{5}-\theta_{0})|0\rangle)|b\rangle+\sin2\theta_{7}\cos2(\theta_{5}-\theta_{0})|0\rangle|c\rangle$\\
$H_{8}$ ($45^\circ$) & \ \ \ \ \ \ \ \ \ \ \ \ \ \ \ \ \ \ \ \ \ \ \ \ \ \ \ \ & $-\cos2\theta_{7}\cos2(\theta_{5}-\theta_{0})|0\rangle|b\rangle+\sin2(\theta_{5}-\theta_{0})|1\rangle|b\rangle$\\
     \hline
     \hline
	\end{tabular}
}
    \label{tab2}
\end{table*}
Since we have done a polarization decomposition of the non-Hermitian operator $B$, we got $-\cos2\theta_{7}\cos2(\theta_{5}-\theta_{0})|0\rangle+\sin2(\theta_{5}-\theta_{0})|1\rangle$, this corresponds to $U_{B}$ and $S_{B}$ in the following matrix form
\begin{eqnarray}
U_{B}=
   \begin{bmatrix}
     \cos2(\theta_{5}-\theta_{0})&\sin2(\theta_{5}-\theta_{0})\\
     \sin2(\theta_{5}-\theta_{0})&-\cos2(\theta_{5}-\theta_{0})
   \end{bmatrix},
S_{B}=
   \begin{bmatrix}
     -\cos2\theta_{7}&0\\
     0&1
   \end{bmatrix}.\label{M2}
\end{eqnarray}

\begin{figure*}
\includegraphics[width=12cm]{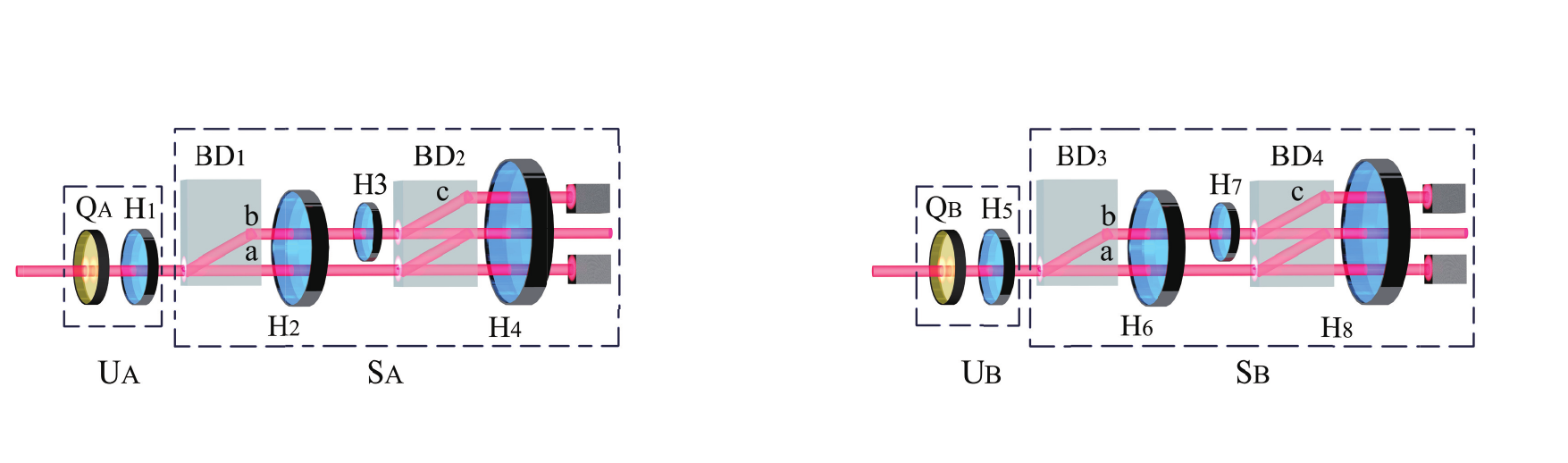}
\caption{Experimental setup of non-Hermitian operators $A=S_{A}U_{A}$ and $B=S_{B}U_{B}$ in the case of complex numbers.} \label{FigIV}
\end{figure*}

The experimental setup of non-Hermitian operators in complex numbers is shown in Fig.~\ref{FigIV}. Again, start with the relevant operation for $A$. In the complex case we make a small change in the treatment of the operator, Table~\ref{tab3} illustrates the relevant processes.
\begin{table*}[htbp]
	\centering
	\caption{The preparation of the non-Hermitian operator $A$ in the complex case.}
\resizebox{\linewidth}{!}{
	\begin{tabular}{cccc}
     \hline
     \hline
		State prepared & Result  \\	
  \hline
$H_{0}$ & $\cos2\theta_{0}|0\rangle+\sin2\theta_{0}|1\rangle$ \\
$Q_{A}$ ($\theta_{A}=0^\circ$) & $\cos2\theta_{0}|0\rangle+i\sin2\theta_{0}|1\rangle$\\
$H_{1}$ & $ (\cos2\theta_{1}\cos2\theta_{0}+i\sin2\theta_{1}\sin2\theta_{0})|0\rangle+(\sin2\theta_{1}\cos2\theta_{0}-i\cos2\theta_{1}\sin2\theta_{0})|1\rangle$\\
$BD_{1}$ & $(\cos2\theta_{1}\cos2\theta_{0}+i\sin2\theta_{1}\sin2\theta_{0})|0\rangle|b\rangle+(\sin2\theta_{1}\cos2\theta_{0}-i\cos2\theta_{1}\sin2\theta_{0})|1\rangle|a\rangle$\\
$H_{2}$ ($45^\circ$) & $(\cos2\theta_{1}\cos2\theta_{0}+i\sin2\theta_{1}\sin2\theta_{0})|1\rangle|b\rangle+(\sin2\theta_{1}\cos2\theta_{0}-i\cos2\theta_{1}\sin2\theta_{0})|0\rangle|a\rangle$\\
$H_{3}$ & $\sin2\theta_{3}(\cos2\theta_{1}\cos2\theta_{0}+i\sin2\theta_{1}\sin2\theta_{0})|0\rangle|b\rangle-\cos2\theta_{3}(\cos2\theta_{1}\cos2\theta_{0}+i\sin2\theta_{1}\sin2\theta_{0})|1\rangle|b\rangle+(\sin2\theta_{1}\cos2\theta_{0}-i\cos2\theta_{1}\sin2\theta_{0})|0\rangle|a\rangle$\\
$BD_{2}$ &  $(-\cos2\theta_{3}(\cos2\theta_{1}\cos2\theta_{0}+i\sin2\theta_{1}\sin2\theta_{0})|1\rangle+(\sin2\theta_{1}\cos2\theta_{0}-i\cos2\theta_{1}\sin2\theta_{0})|0\rangle)|b\rangle+\sin2\theta_{3}(\cos2\theta_{1}\cos2\theta_{0}+i\sin2\theta_{1}\sin2\theta_{0})|0\rangle|c\rangle$\\
$H_{4}$ ($45^\circ$) & $-\cos2\theta_{3}(\cos2\theta_{1}\cos2\theta_{0}+i\sin2\theta_{1}\sin2\theta_{0})|0\rangle|b\rangle+(\sin2\theta_{1}\cos2\theta_{0}-i\cos2\theta_{1}\sin2\theta_{0})|1\rangle|b\rangle$\\
     \hline
     \hline
	\end{tabular}
}
    \label{tab3}
\end{table*}

Since we have done a polarization decomposition of the non-Hermitian operator $A$ in the complex case, we got $-\cos2\theta_{3}(\cos2\theta_{1}\cos2\theta_{0}+i\sin2\theta_{1}\sin2\theta_{0})|0\rangle+(\sin2\theta_{1}\cos2\theta_{0}-i\cos2\theta_{1}\sin2\theta_{0})|1\rangle$, this corresponds to $U_{A}$ and $S_{A}$ in the following matrix form

\begin{eqnarray}
U_{A}=
   \begin{bmatrix}
     \cos2(\theta_{1}-\theta_{0})&i\sin2(\theta_{1}-\theta_{0})\\
     \sin2(\theta_{1}-\theta_{0})&-i\cos2(\theta_{1}-\theta_{0})
   \end{bmatrix},
S_{A}=
   \begin{bmatrix}
     -\cos2\theta_{3}&0\\
     0&1
   \end{bmatrix}.\label{M3}
\end{eqnarray}
The experimental setup of the non-Hermitian operator $B$ in complex numbers is shown in Fig.~\ref{FigIV}. The details of the process implementation can be found in Table~\ref{tab4}.
\begin{table*}[htbp]
	\centering
	\caption{The preparation of the non-Hermitian operator $B$ in the complex case.}
\resizebox{\linewidth}{!}{
	\begin{tabular}{cccc}
     \hline
     \hline
		State prepared & Result  \\	
  \hline
$H_{0}$ & $\cos2\theta_{0}|0\rangle+\sin2\theta_{0}|1\rangle$ \\
$Q_{B}$ ($\theta_{B}=0^\circ$) & $\cos2\theta_{0}|0\rangle+i\sin2\theta_{0}|1\rangle$\\
$H_{5}$ & $ (\cos2\theta_{5}\cos2\theta_{0}+i\sin2\theta_{5}\sin2\theta_{0})|0\rangle+(\sin2\theta_{5}\cos2\theta_{0}-i\cos2\theta_{5}\sin2\theta_{0})|1\rangle$\\
$BD_{3}$ & $(\cos2\theta_{5}\cos2\theta_{0}+i\sin2\theta_{5}\sin2\theta_{0})|0\rangle|b\rangle+(\sin2\theta_{5}\cos2\theta_{0}-i\cos2\theta_{5}\sin2\theta_{0})|1\rangle|a\rangle$\\
$H_{6}$ ($45^\circ$) & $(\cos2\theta_{5}\cos2\theta_{0}+i\sin2\theta_{5}\sin2\theta_{0})|1\rangle|b\rangle+(\sin2\theta_{5}\cos2\theta_{0}-i\cos2\theta_{5}\sin2\theta_{0})|0\rangle|a\rangle$\\
$H_{7}$ & $\sin2\theta_{7}(\cos2\theta_{5}\cos2\theta_{0}+i\sin2\theta_{5}\sin2\theta_{0})|0\rangle|b\rangle-\cos2\theta_{7}(\cos2\theta_{5}\cos2\theta_{0}+i\sin2\theta_{5}\sin2\theta_{0})|1\rangle|b\rangle+(\sin2\theta_{5}\cos2\theta_{0}-i\cos2\theta_{5}\sin2\theta_{0})|0\rangle|a\rangle$\\
$BD_{4}$ &  $(-\cos2\theta_{7}(\cos2\theta_{5}\cos2\theta_{0}+i\sin2\theta_{5}\sin2\theta_{0})|1\rangle+(\sin2\theta_{5}\cos2\theta_{0}-i\cos2\theta_{5}\sin2\theta_{0})|0\rangle)|b\rangle+\sin2\theta_{7}(\cos2\theta_{5}\cos2\theta_{0}+i\sin2\theta_{5}\sin2\theta_{0})|0\rangle|c\rangle$\\
$H_{8}$ ($45^\circ$) & $-\cos2\theta_{7}(\cos2\theta_{5}\cos2\theta_{0}+i\sin2\theta_{5}\sin2\theta_{0})|0\rangle|b\rangle+(\sin2\theta_{5}\cos2\theta_{0}-i\cos2\theta_{5}\sin2\theta_{0})|1\rangle|b\rangle$\\
     \hline
     \hline
	\end{tabular}
}
    \label{tab4}
\end{table*}
Since we have done a polarization decomposition of the non-Hermitian operator $B$, we got $-\cos2\theta_{7}(\cos2\theta_{5}\cos2\theta_{0}+i\sin2\theta_{5}\sin2\theta_{0})|0\rangle+(\sin2\theta_{5}\cos2\theta_{0}-i\cos2\theta_{5}\sin2\theta_{0})|1\rangle$, this corresponds to $U_{B}$ and $S_{B}$ in the following matrix form

\begin{eqnarray}
U_{B}=
   \begin{bmatrix}
     \cos2(\theta_{5}-\theta_{0})&i\sin2(\theta_{5}-\theta_{0})\\
     \sin2(\theta_{5}-\theta_{0})&-i\cos2(\theta_{5}-\theta_{0})
   \end{bmatrix},
S_{B}=
   \begin{bmatrix}
     -\cos2\theta_{7}&0\\
     0&1
   \end{bmatrix}.\label{M4}
\end{eqnarray}

\begin{figure*}
\includegraphics[width=12cm]{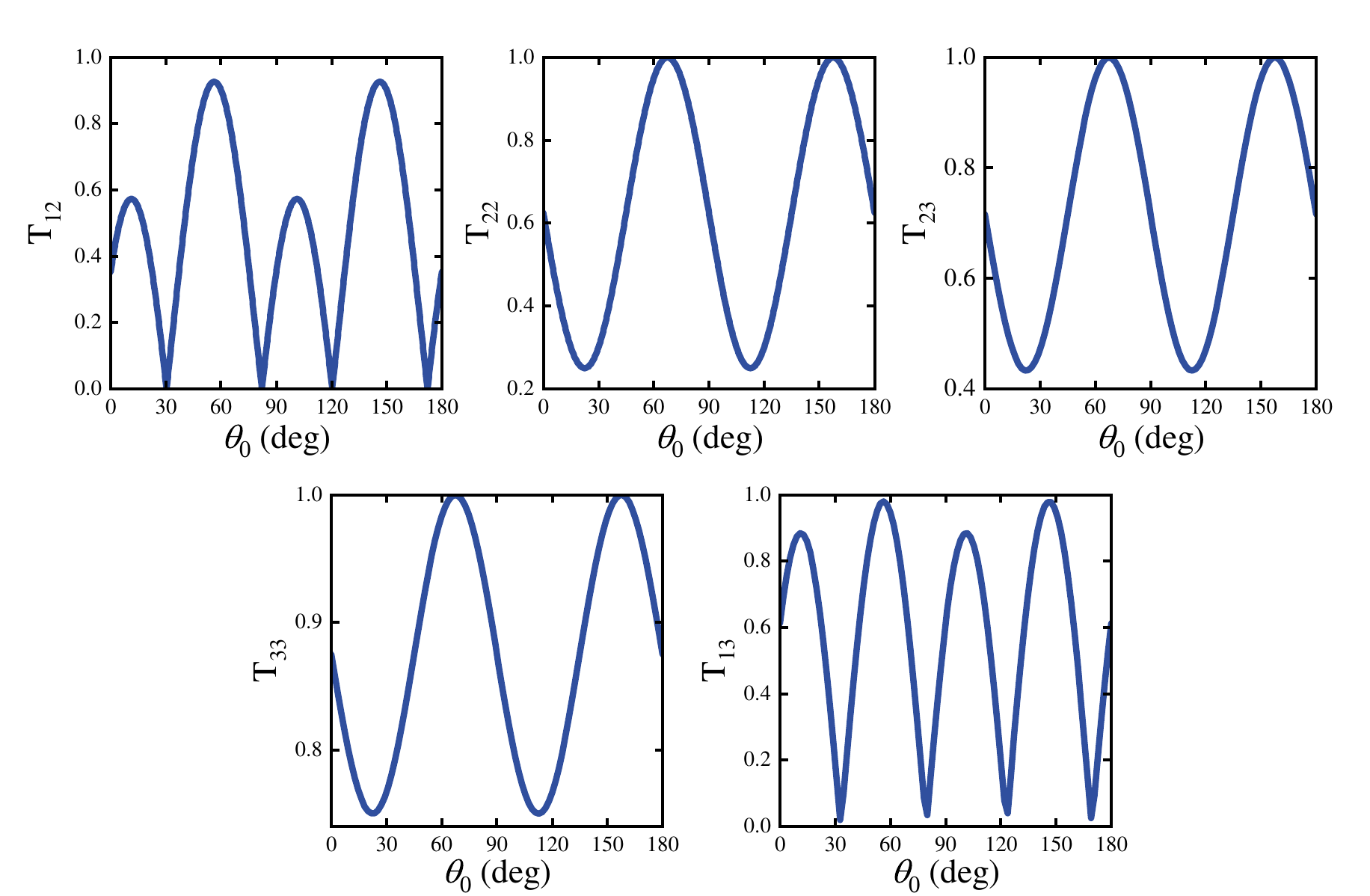}
\caption{The theoretical values for some items in an uncertain relation, non-Hermitian operators are real. (a) $T_{12}=|\langle A^\dag\rangle|$. (b)$T_{22}=|\langle A^\dag A\rangle|$. (c) $T_{23}=|\langle A^\dag B\rangle|$. (d) $T_{33}=|\langle B^\dag B\rangle|$. (e) $T_{13}=|\langle B\rangle|$.} \label{FigV}
\end{figure*}

\begin{figure*}
\includegraphics[width=11cm]{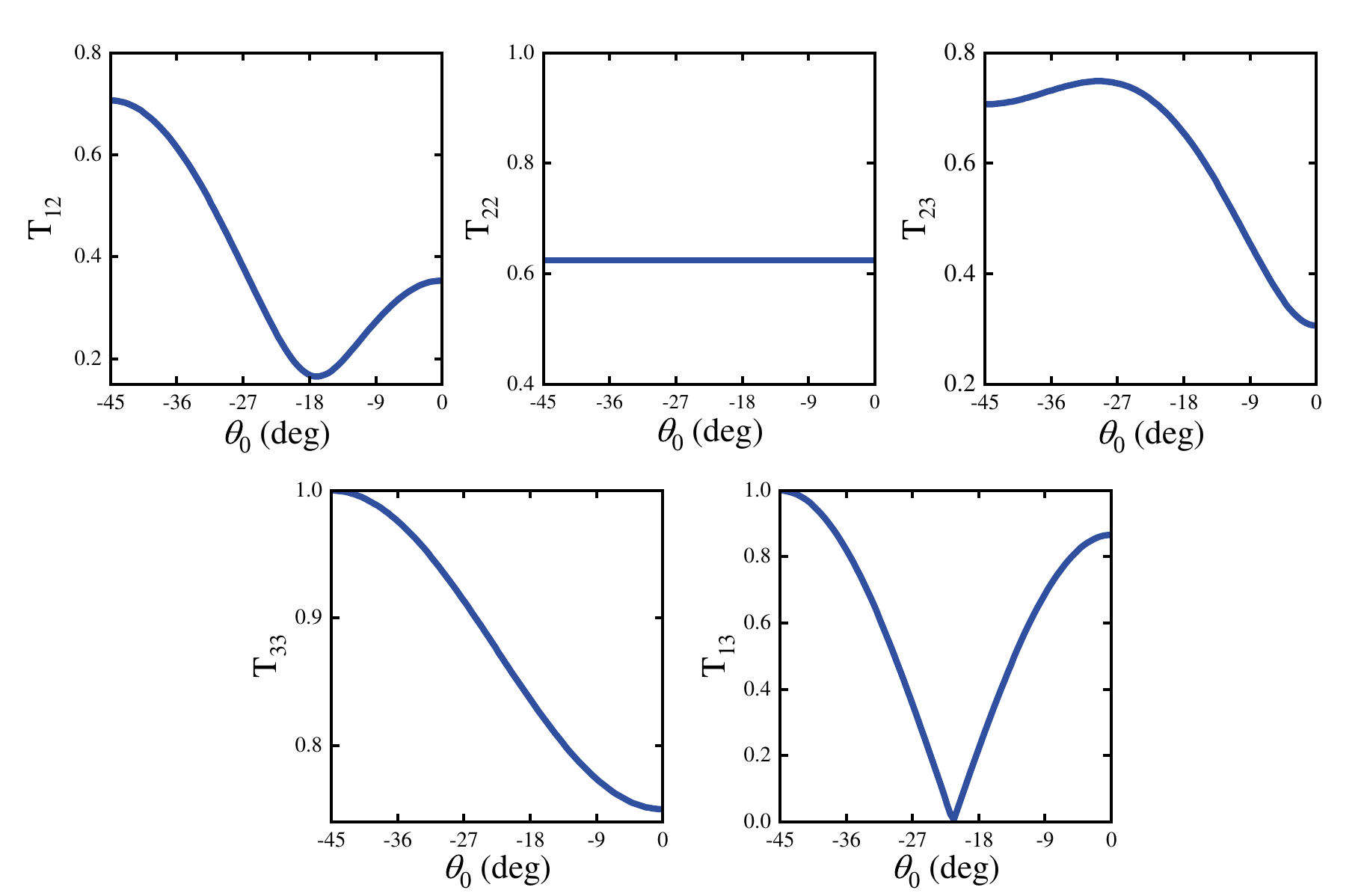}
\caption{The theoretical values for some items in an uncertain relation, non-Hermitian operators are complex. (a) $T_{12}=|\langle A^\dag\rangle|$. (b)$T_{22}=|\langle A^\dag A\rangle|$. (c) $T_{23}=|\langle A^\dag B\rangle|$. (d) $T_{33}=|\langle B^\dag B\rangle|$. (e) $T_{13}=|\langle B\rangle|$.} \label{FigVI}
\end{figure*}

\begin{figure*}[h]
\includegraphics[width=12cm]{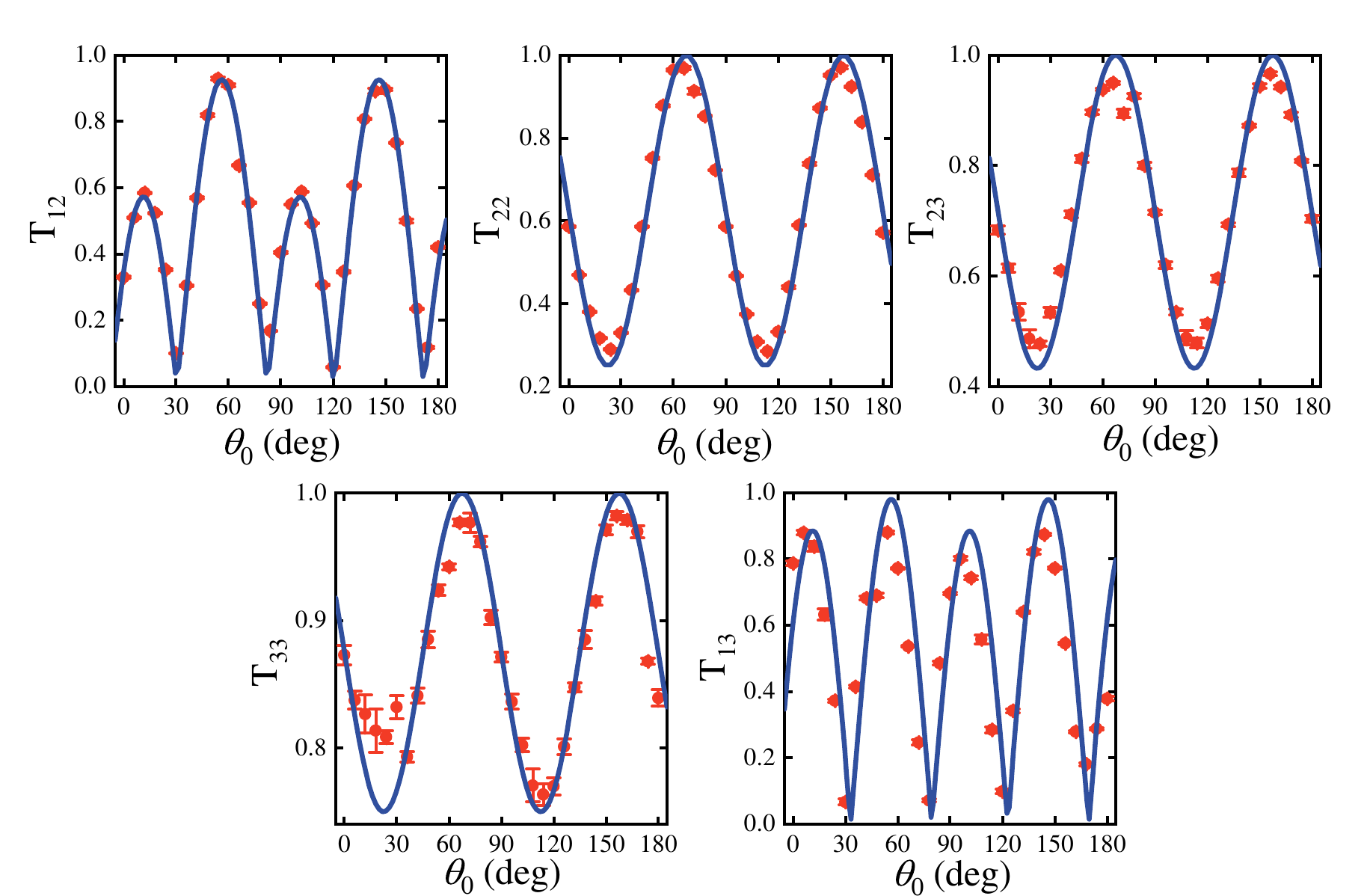}
\caption{The experimental data results compared with theoretical values for some items in an uncertain relation, non-Hermitian operators are real. (a) $T_{22}=|\langle A^\dag A\rangle|$. (b) $T_{13}=|\langle B\rangle|$. (c) $T_{12}=|\langle A^\dag\rangle|$. (d) $T_{23}=|\langle A^\dag B\rangle|$. (e) $T_{33}=|\langle B^\dag B\rangle|$. The error bar is calculated by the error propagation formula.} \label{FigVIII}
\end{figure*}

\begin{figure*}[h]
\includegraphics[width=12cm]{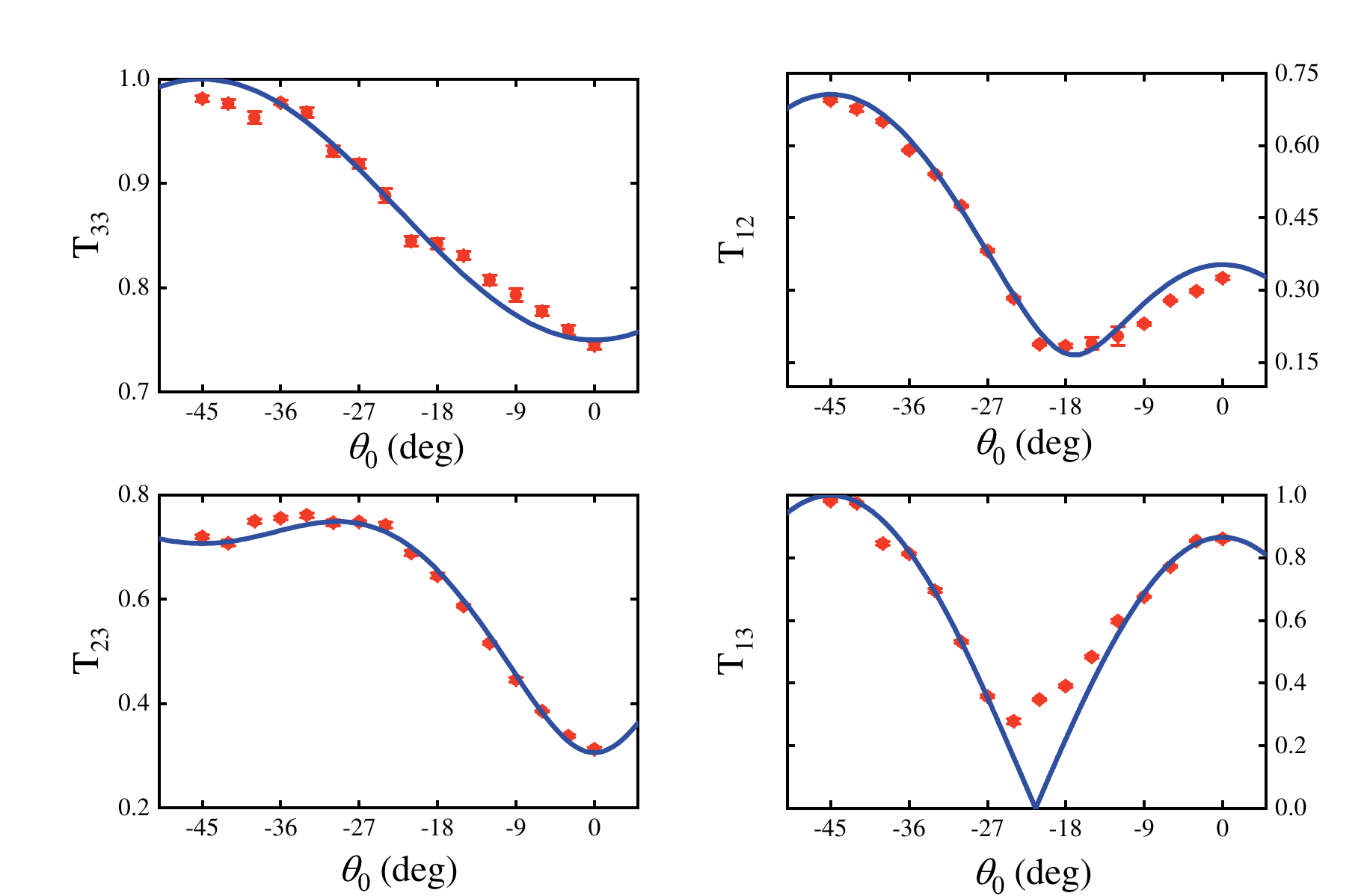}
\caption{The experimental data results compared with theoretical values for some items in an uncertain relation, non-Hermitian operators are complex. (a) $T_{33}=|\langle B^\dag B\rangle|$. (b)$T_{12}=|\langle A^\dag\rangle|$. (c) $T_{23}=|\langle A^\dag B\rangle|$. (d) $T_{13}=|\langle B\rangle|$. The error bar is calculated by the error propagation formula.} \label{FigIX}
\end{figure*}

\subsection{The theoretical value of the uncertainty relation of non-Hermitian operators}
Here, we illustrate the theoretical result of the uncertainty relation of non-Hermitian operators. As shown in Fig.~\ref{FigV}, the non-Hermitian operator $A$, $B$ is the real case. $\theta_{1}=22.5^\circ$, $\theta_{3}=60^\circ$, $\theta_{5}=22.5^\circ$, $\theta_{7}=75^\circ$. The theoretical values we calculated are as follows
\begin{eqnarray}
|T_{12}|&=&\dfrac{\left|-\dfrac{1}{2}+\dfrac{3}{2}\Big(\cos{\dfrac{\theta_{0}}{45}}\pi+\sin{\dfrac{\theta_{0}}{45}}\pi\Big)\right|}{2\sqrt{2}},\\
|T_{13}|&=&\dfrac{\left|\Bigg(-1+\dfrac{\sqrt{3}}{2}\Bigg)+\Bigg(1+\dfrac{\sqrt{3}}{2}\Bigg)\Bigg(\cos{\dfrac{\theta_{0}}{45}}\pi+\sin{\dfrac{\theta_{0}}{45}}\pi\Bigg)\right|}{2\sqrt{2}},\\
|T_{22}|&=&\dfrac{1}{2}\left|\dfrac{5}{4}-\dfrac{3}{8}\sin{\dfrac{\theta_{0}}{45}}\pi\right|,\\
|T_{33}|&=&\dfrac{1}{2}\left|\dfrac{7}{4}+\Bigg(-1+\dfrac{\sqrt{3}}{2}\Bigg)\Bigg(1+\dfrac{\sqrt{3}}{2}\Bigg)\sin{\dfrac{\theta_{0}}{45}}\pi\right|,\\
|T_{23}|&=&\dfrac{1}{2}\left|1+\dfrac{\sqrt{3}}{4}+\Bigg(-1+\dfrac{\sqrt{3}}{4}\Bigg)\sin{\dfrac{\theta_{0}}{45}}\pi\right|.
\end{eqnarray}

There are some differences between the case where the non-Hermitian operator is complex and the case of real numbers, and its theoretical curve is shown in Fig.~\ref{FigVI}, $\theta_{1}=22.5^\circ$, $\theta_{3}=60^\circ$, $\theta_{5}=0^\circ$, $\theta_{7}=75^\circ$, $\theta_{A}=0^\circ$ and $\theta_{B}=0^\circ$. In the same way, we also give the theoretical values corresponding to the initial state angle $\theta_{0}$
\begin{eqnarray}
|T_{12}|&=&\dfrac{-1+3\cos{\dfrac{\theta_{0}}{45}}\pi+i\sin{\dfrac{\theta_{0}}{45}}\pi}{4\sqrt{2}},\\
|T_{13}|&=&\dfrac{\sqrt{3}}{2}\Bigg(\cos{\dfrac{\theta_{0}}{90}}\pi\Bigg)^2+\Bigg(\sin{\dfrac{\theta_{0}}{90}}\pi\Bigg)^2,\\
|T_{22}|&=&\dfrac{5}{8},\\
|T_{33}|&=&\left|\dfrac{3}{4}(\cos{\dfrac{\theta_{0}}{45}}\pi)^2+(\sin{\dfrac{\theta_{0}}{45}}\pi)^2\right|,\\
|T_{23}|&=&\dfrac{\left|4+\sqrt{3}+\sqrt{3}\Bigg(\cos{\dfrac{\theta_{0}}{45}}\pi-i\sin{\dfrac{\theta_{0}}{45}}\pi\Bigg)-4\Bigg(\cos{\dfrac{\theta_{0}}{45}}\pi+i\sin{\dfrac{\theta_{0}}{45}}\pi\Bigg)\right|}{8\sqrt{2}}.
\end{eqnarray}

\section*{Appendix C. The result of the experiment}

We present a graph comparing theoretical values of $T_{22}=|\langle A^\dag A\rangle|$, $T_{13}=|\langle B\rangle|$, $T_{12}=|\langle A^\dag\rangle|$, $T_{23}=|\langle A^\dag B\rangle|$ and $T_{33}=|\langle B^\dag B\rangle|$ with experimental results. Fig.~\ref{FigVIII} shows the real case and Fig.~\ref{FigIX} shows the case of complex numbers. Since $T_{22}=|\langle A^\dag A\rangle|$ in the complex case is a constant, we don't measure it. The experimental results are acceptable within the error bar. As shown in Fig.~\ref{FigVIII} for $T_{33}$ and Fig.~\ref{FigIX} for $T_{13}$, the numerical results for this point may be due to a lack of precision in the angle of the waveplate or a reduction in the interference of the Sagnac ring at the time of measurement, preventing it from reaching the lowest point corresponding to the theoretical value.

\section*{Appendix D. Application for entanglement detection}
The uncertainty relations for non-Hermitian operators can also be applied to entanglement detection, similar to Hermitian operator uncertainty relations in \cite{Hofm}. 
In this section, we will present the necessary derivations.

A quantum channel $\mathcal{E}$ with $\left\lbrace E_{k}\right\rbrace$ being the chosen Kraus operators, can be expressed as follows
\begin{eqnarray}
\mathcal{E}(\rho)=\sum_{k}E_{k}\rho E_{k}^\dag,\label{E1}
\end{eqnarray}
with $\sum_{k}E_{k}^\dag E_{k}=\mathbb{I}$. The variance of $E_{k}$ for a pure state $|\psi\rangle$ is 
\begin{eqnarray}
\Delta E_{k}^2=\langle\psi|E_{k}^\dag E_{k}|\psi\rangle-\langle\psi|E_{k}^\dag|\psi\rangle\langle\psi|E_{k}|\psi\rangle,\label{E2}
\end{eqnarray}
then we can use a relationship between the quantum channel fidelity $F$ and the sum of variances, which has been first time derived in  \cite{Pati}, 
\begin{eqnarray}
\sum_{k}\Delta E_{k}^2=1-\sum_{k}\langle\psi|E_{k}^\dag|\psi\rangle\langle\psi|E_{k}|\psi\rangle=1-\langle\psi|\mathcal{E}(|\psi\rangle\langle\psi|)|\psi\rangle=1-F\ge 1-F_{max}(\mathcal{E}),\label{E3}
\end{eqnarray}
where
\begin{eqnarray}
F_{max}(\mathcal{E})=\max_{|\phi\rangle} \ \langle\phi|\mathcal{E}(|\phi\rangle\langle\phi|)|\phi\rangle.\label{E4}
\end{eqnarray}
Here we do not show the analytical solution form of the maximum quantum channel fidelity $F_{max}(\mathcal{E})$, but relevant examples of $F_{max}(\mathcal{E})$ have been given in Refs.~\cite{Erns,Chel,Siud}, which can provide reference for subsequent researchers.

Consider a bipartite sytem, if a state $\mathcal{\rho}_{\rm{AB}}$ is separable, i.e. it can be written as
\begin{eqnarray}
\mathcal{\rho}_{\rm{AB}}=\sum_{i}p_{i}|\psi_{i}^{\rm{A}}\rangle\langle\psi_{i}^{\rm{A}}|\otimes|\psi_{i}^{\rm{B}}\rangle\langle\psi_{i}^{\rm{B}}|.\label{E5}
\end{eqnarray}
For each subsystem $\rm{A}$ and $\rm{B}$, there exist a quantum channel $\mathcal{E}^{\rm{A}}$ and $\mathcal{E}^{\rm{B}}$. Based on Eq.~(\ref{E3}), we can get respectively
\begin{eqnarray}
\sum_{k}\Delta (E_{k}^{\rm{A}})^2\ge 1-F_{max}(\mathcal{E}^{\rm{A}}),\label{SS1} \\
\sum_{k}\Delta (E_{k}^{\rm{B}})^2\ge 1-F_{max}(\mathcal{E}^{\rm{B}}).\label{SS2}
\end{eqnarray}
For a mixed state $\rho=\sum_i p_i |\psi_i\rangle\langle\psi_i|$, the variance is a concave function, i.e.
\begin{eqnarray}
\Delta(M)_{\rho}^2\ge\sum_{i}p_{i}\Delta(M)_{|\psi_{i}\rangle}^2. 
\end{eqnarray}
Thus, for a separable state $\rho_{\rm{AB}}$ in Eq.~(\ref{E5}) we have
\begin{eqnarray}
\sum_{k}\Delta(E_{k}^{\rm{A}}\otimes\mathbb{I}+\mathbb{I}\otimes E_{k}^{\rm{B}})_{\mathcal{\rho}_{\rm{AB}}}^2 &&\ge\sum_{k}\sum_{i}p_{i}\Delta(E_{k}^{\rm{A}}\otimes\mathbb{I}+\mathbb{I}\otimes E_{k}^{\rm{B}})_{|\psi_{i}^{\rm{A}}\rangle\otimes|\psi_{i}^{\rm{B}}\rangle}^2\nonumber\\&&=\sum_{i}p_{i}\sum_{k}\left[\Delta(E_{k}^{\rm{A}})_{|\psi_{i}^{\rm{A}}\rangle}^2+\Delta(E_{k}^{\rm{B}})_{|\psi_{i}^{\rm{B}}\rangle}^2\right].\label{E6}
\end{eqnarray}

It is obviously, we can get
\begin{eqnarray}
\sum_{i}p_{i}\sum_{k}\left[\Delta(E_{k}^{\rm{A}})_{|\psi_{i}^{\rm{A}}\rangle}^2+\Delta(E_{k}^{\rm{B}})_{|\psi_{i}^{\rm{B}}\rangle}^2\right] \ge 1-F_{max}(\mathcal{E}^{\rm{A}})+1-F_{max}(\mathcal{E}^{\rm{B}}).
\end{eqnarray}
For each $\Delta(E_{k}^{\rm{A}})_{|\psi_{i}^{\rm{A}}\rangle}^2$ and $\Delta(E_{k}^{\rm{B}})_{|\psi_{i}^{\rm{B}}\rangle}^2$, the Eqs.~(\ref{SS1}) and (\ref{SS2}) always hold with $\sum_{i}p_{i}=1$.
Then we have the final expression as 
\begin{eqnarray}
\sum_{k}\Delta(E_{k}^{\rm{A}}\otimes\mathbb{I}+\mathbb{I}\otimes E_{k}^{\rm{B}})_{\mathcal{\rho}_{\rm{AB}}}^2\ge 2-F_{max}(\mathcal{E}^{\rm{A}})-F_{max}(\mathcal{E}^{\rm{B}}),\label{E7}
\end{eqnarray}
this inequality must hold for all separable states. If Eq.~(\ref{E7}) is violated for a quantum state, it must be entangled.

\end{document}